\documentclass[aps,prb,reprint,showpacs]{revtex4-1}
\usepackage[dvipdfmx]{graphicx}
\usepackage{subfigmat}
\usepackage{amsmath,amssymb}
\usepackage{color}
\usepackage{bm}
\usepackage{hyperref}
\usepackage{comment}

\usepackage{cancel}

\usepackage{upgreek}

\newcommand{\tempinv}{\upbeta}

\begin{document}

\title{Effect of vertex corrections on the enhancement of Gilbert damping in spin pumping \\ into a two-dimensional electron gas}

\author{M. Yama,$^{1}$ M. Matsuo,$^{2,3,4,5}$ T. Kato$^{1}$,}
\affiliation{
${^1}$Institute for Solid State Physics, The University of Tokyo, Kashiwa, Japan\\
${^2}$Kavli Institute for Theoretical Sciences, University of Chinese Academy of Sciences, Beijing, China\\
${^3}$CAS Center for Excellence in Topological Quantum Computation, University of Chinese Academy of Sciences, Beijing, China\\
${^4}$Advanced Science Research Center, Japan Atomic Energy Agency, Tokai, Japan\\
${^5}$RIKEN Center for Emergent Matter Science (CEMS), Wako, Saitama, Japan\\
}

\date{\today}

\begin{abstract}
We theoretically consider the effect of vertex correction on spin pumping from a ferromagnetic insulator (FI) into a two-dimensional electron gas (2DEG) in which the Rashba and Dresselhaus spin-orbit interactions coexist. 
The Gilbert damping in the FI is enhanced by elastic spin-flipping or magnon absorption.
We show that the Gilbert damping due to elastic spin-flipping is strongly enhanced by the vertex correction when the ratio of the two spin-orbit interactions is near a special value at which the spin relaxation time diverges while that due to magnon absorption shows only small modification.
We also show that the shift in the resonant frequency due to elastic spin-flipping is strongly enhanced in a similar way as the Gilbert damping.
\end{abstract}
\maketitle 

\section{Introduction}
\label{sec:introduction}

In the field of spintronics~\cite{Zutic2004,Tsymbal2021}, spin pumping has long been used as a method of injecting spins into various materials~\cite{Tserkovnyak2002,Tserkovnyak2005,Hellman2017}.
Spin pumping was first employed to inject spins from a ferromagnetic metal into an adjacent normal metal (NM)~\cite{Mizukami2001,Mizukami2002,Saitoh2006,Ando2008}. Subsequently, it was used on ferromagnetic insulator (FI)/NM junctions~\cite{Kajiwara2010}.
Because spin injection is generally related to the loss of the magnetization in ferromagnets, it affects the Gilbert damping measured in ferromagnetic resonance (FMR) experiments~\cite{han2020spin}.
When we employ spin injection from the FI, the modulation of the Gilbert damping reflects the properties of the spin excitation in the adjacent materials, such as magnetic thin films~\cite{qiu2016spin}, magnetic impurities on metal surfaces~\cite{yamamoto2021}, and superconductors~\cite{inoueSpinPumpingSuperconductors2017,Kato2020,Ominato2021,Ominato2022}.
This is in clear contrast with the Gilbert damping of a bulk FI, which reflects properties of electrons and phonons~\cite{Garate2009a,Garate2009b,Liu2017}.

An attractive strategy is to combine spin pumping with spin-related transport phenomena in semiconductor microstructures~\cite{Zutic2004,Awschalom2007}.
A two-dimensional electron gas (2DEG) in a semiconductor heterostructure is an easily controlled physical system that has been used in spintronics devices~\cite{Datta1990,Srisongmuang2008,Akabori2012,Feng2017}. A 2DEG system has two types of spin-orbit interaction, i.e., Rashba~\cite{Bychkov1984,Rashba2015} and Dresselhaus spin-orbit interactions~\cite{Dresselhaus1955,Rocca1988}. 

\begin{figure}[tb]
\centering
\includegraphics[width=80mm]{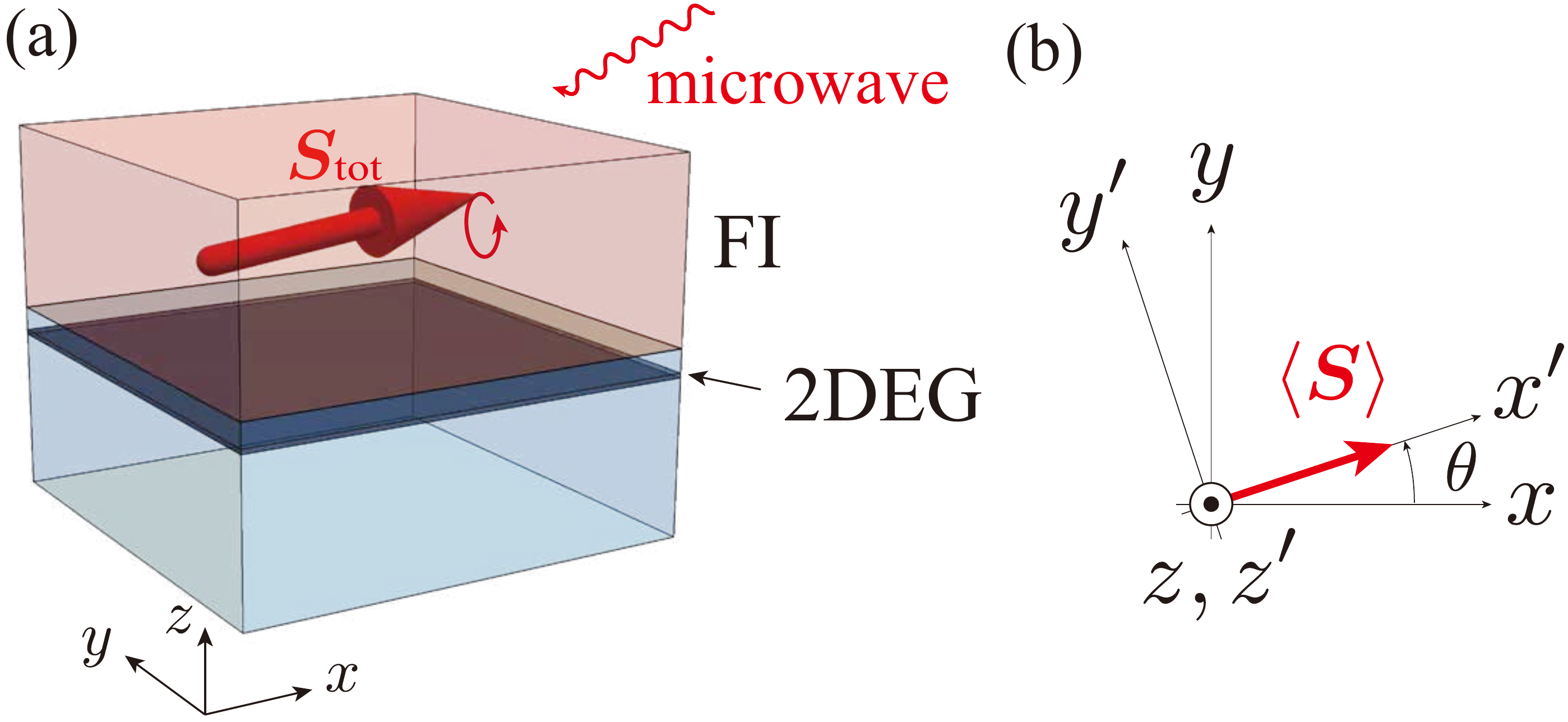}
\caption{(a) Schematic picture of junction composed of a ferromagnetic insulator (FI) and a two-dimensional electron gas (2DEG) realized in a semiconductor heterostructure.
$\bm{S}_{\rm tot}$ indicates the total spin of the FI.
We consider a uniform spin precession of the FI induced by microwave irradiation.
(b) Laboratory coordinates $(x,y,z)$ and the magnetization-fixed coordinates $(x',y',z')$. The red arrow indicates the expectation value of the spontaneous spin polarization of the FI, $\langle \bm{S}\rangle$.}
\label{fig:setup}
\end{figure}

In our previous work~\cite{Yama2021}, we theoretically studied spin pumping into a 2DEG in semiconductor heterostructures with both Rashba and Dresselhaus spin-orbit interactions, which can be regarded as a prototype for a 2DEG with a complex spin-texture near the Fermi surface [see Fig.~\ref{fig:setup}~(a)].
In that study, we formulated the modulation of the Gilbert damping in the FI by using the second-order perturbation with respect to the interfacial  coupling~\cite{Ohnuma2014,Matsuo2018,Kato2019,Kato2020,Ominato2020a,Ominato2020b} and related it to the dynamic spin susceptibility of the 2DEG. We further calculated the spin susceptibility and obtained characteristic features of the Gilbert damping modulation. This modulation contains two contributions: elastic spin-flipping, which dominates at low resonant frequencies, and magnon absorption, which dominates at high resonant frequencies. 
In addition, we clarified that these contributions have different dependence on the in-plane azimuth angle $\theta$ of the ordered spin in the FI [see Fig.~\ref{fig:setup}~(b)].

When the Rashba and Dresselhaus spin-orbit interactions have almost equal magnitudes, spin relaxation by nonmagnetic impurity scattering is strongly suppressed because the direction of the effective Zeeman field generated by the spin-orbit interactions is unchanged along the Fermi surface.
Due to this substantial suppression of spin relaxation, there emerge characteristic physical phenomena such as the persistent spin helix state~\cite{Bernevig2006,Weber2007,Koralek2009,Sasaki2014}.
In general, the vertex corrections have to be taken into account to treat various conservation laws, i.e., the charge, spin, momentum, and energy conservation laws in calculation of the response functions\cite{Baym1961,Baym1962,Bruus2004,Akkermans2007}.
Therefore, for better description of realistic systems, we need to consider vertex correction, which captures effect of impurity more accurately by reflecting conservation laws.
However, the vertex corrections were neglected in our previous work~\cite{Yama2021}.
This means that our previous calculation should fail when the Rashba and Dresselhaus spin-orbit interactions compete.

In this study, we consider the same setting, i.e., a junction composed of an FI and a 2DEG as shown in Fig.~\ref{fig:setup}~(a), and examine effect of the spin conservation law by taking the vertex correction into account.
We theoretically calculate the modulation of the Gilbert damping and the shift in the FMR frequency by solving the Bethe-Salpeter equation within the ladder approximation.
We show that the vertex correction substantially changes the results, in particular, when the strengths of the Rashba- and Dresselhaus-type spin-orbit interactions are chosen to be almost equal but slightly different; Specifically, both the Gilbert damping and the FMR frequency shift are largely enhanced at low resonant frequencies reflecting strong suppression of spin relaxation.
This remarkable feature should be able to be observed experimentally.
In contrast, the vertex correction changes their magnitude only slightly at high resonant frequencies.

Before describing our calculation, we briefly comment on study of the vertex corrections in a different context.
In early studies of the spin Hall effect, there was a debate on the existence of intrinsic spin Hall effect~\cite{Sinova2015,Murakami2003,Sinova2004}.
By considering the vertex corrections, the spin Hall conductivity, which is calculated from the correlation function between the current and spin current, vanishes in the presence of short-range disorder for simple models even if its strength is infinitesimally small\cite{Inoue2004,Dimitrova2005,Raimondi2012}.
This seemingly contradictory result stimulated theoretical researches on realistic modified models~\cite{Murakami2004,Krotkov2006} as well as definition of the spin current~\cite{Shi2006,Zhang2008,Gorini2012,Tatara2018,Shitade2022}.
However, we stress that the vertex corrections for the dynamic spin susceptibility, which is calculated from the spin-spin correlation function, have no such subtle problem\cite{Fulde1968} because it does not include the spin current.

The rest of this work is organized as follows. In Sec.~\ref{sec:formulation}, we briefly summarize our model of the FI/2DEG junction and describe a general formulation for the magnon self-energy following Ref.~\onlinecite{Yama2021}.
In Sec.~\ref{sec:VC}, we formulate the vertex correction that corresponds to the self-energy in the Born approximation.
We show the modulation of the Gilbert damping and the shift in the FMR frequency in Secs.~\ref{sec:effGIL} and \ref{sec:shift}, respectively, and discuss the effect of the vertex correction in detail. 
Finally, we summarize our results in Sec.~\ref{sec:summary}. The six Appendices detail the calculation in Sec.~\ref{sec:VC}.

\section{Formulation}
\label{sec:formulation}

Here, we describe a model for the FI/2DEG junction shown in Fig.~\ref{fig:setup}~(a) and formulate the spin relaxation rate in an FMR experiment.
Because we have already given a detailed formulation on this model in our previous paper~\cite{Yama2021}, we will briefly summarize it here.

\begin{figure}[tb]
\centering
\includegraphics[width=75mm]{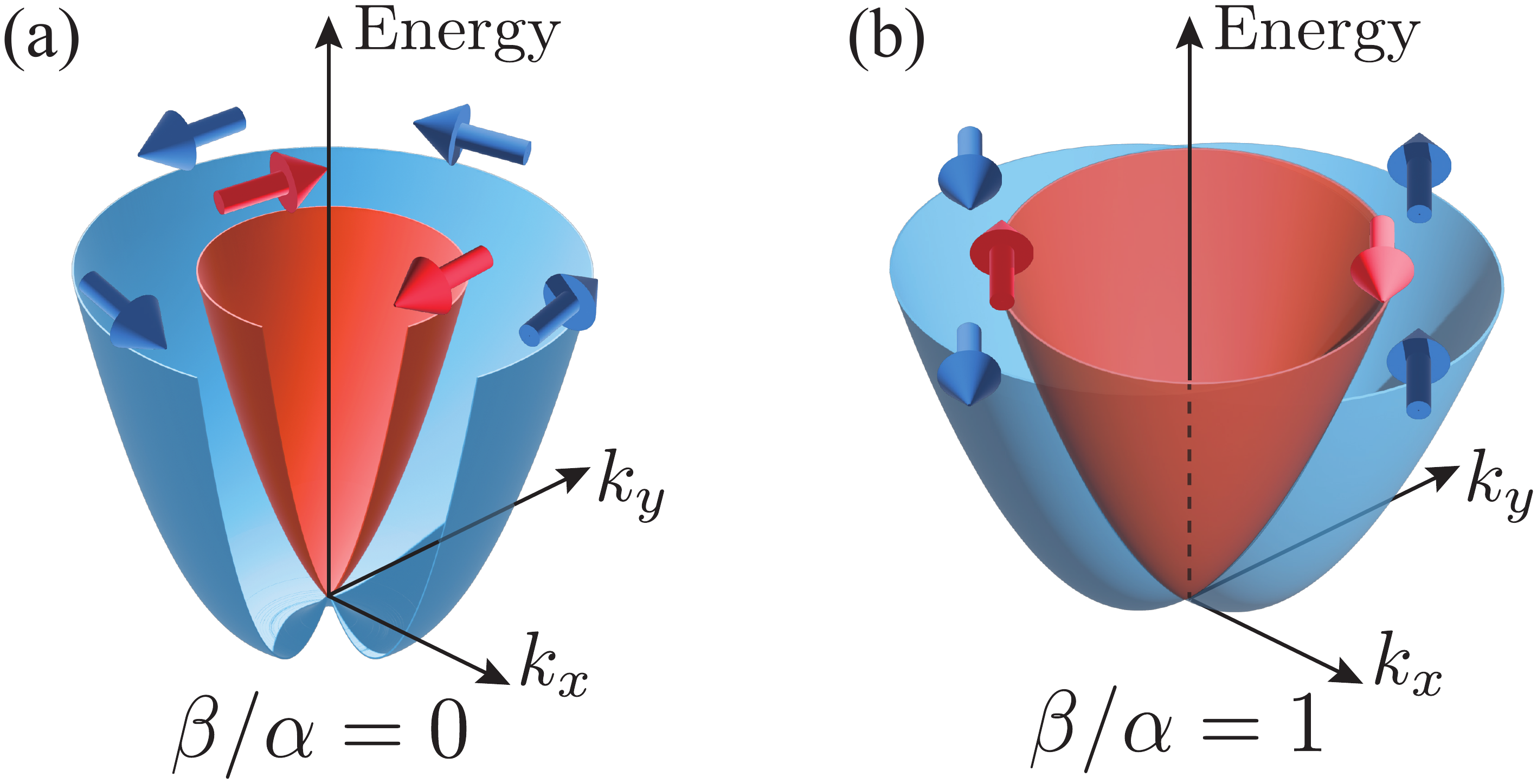}
\caption{Schematic picture of the spin-splitting energy bands of 2DEG for (a) $\beta/\alpha=0$ and (b) $\beta/\alpha=1$. The red and blue arrows represent spin polarization of each band. 
In the case of (b), the spin component in the direction of the azimuth angle $3\pi/4$ is conserved.
}
\label{fig:band}
\end{figure}

\subsection{Two-dimensional electron gas}
\label{sec:model}

We consider a 2DEG whose Hamiltonian is given as $H_{\rm NM}=H_{\rm kin} + H_{\rm imp}$, where $H_{\rm kin}$ and $H_{\rm imp}$ describe the kinetic energy and the impurity, respectively.
The kinetic energy is given as
\begin{align}
H_{\rm kin} &= \sum_{\bm k} (c_{{\bm k}\uparrow} ^\dagger \ c_{{\bm k}\downarrow}^\dagger) \, \hat{h}_{\bm k} \Bigl( \begin{array}{cc} c_{{\bm k}\uparrow} \\ c_{{\bm k}\downarrow} \end{array} \Bigr), \\
\hat{h}_{\bm k} &= \xi_{\bm k} \hat{I} - {\bm h}_{\rm eff}(\bm{k}) \cdot {\bm \sigma} ,
\end{align}
where $c_{{\bm k}\sigma}$ is the annihilation operator of conduction electrons with wave number ${\bm k}=(k_x,k_y)$ and $z$ component of the spin, $\sigma$ ($=\uparrow, \downarrow$), $\hat{I}$ is a $2\times 2$ identity matrix, $\sigma_a$ ($a=x,y,z$) are the Pauli matrices, $\xi_{\bm k}=\hbar^2 {\bm k}^2/2m^{*}-\mu$ is the kinetic energy measured from the chemical potential, and $m^*$ is an effective mass. Hereafter, we assume that the Fermi energy is much larger than the other energy scales such as the spin-orbit interactions, the temperature, and the ferromagnetic resonance energy.
Then, the low-energy part of the spin susceptibility depends on the chemical potential $\mu$ and the effective mass $m^*$ only through the density of states at the Fermi energy, $D(\epsilon_{\rm F})$.

The spin-orbit interaction is described by the effective Zeeman field,
\begin{align}
&{\bm h}_{\rm eff}(\bm{k}) = |{\bm k}|(-\alpha \sin \varphi - \beta \cos \varphi, \alpha \cos \varphi + \beta \sin \varphi, 0) \nonumber \\
&\simeq k_{\rm F}(-\alpha \sin \varphi - \beta \cos \varphi, \alpha \cos \varphi + \beta \sin \varphi, 0), \label{eq:heffkf}
\end{align}
where $\alpha$ and $\beta$ respectively denote the amplitudes of the Rashba- and Dresselhaus-type spin-orbit interactions and the electron wave number is expressed by polar coordinates  as $(k_x,k_y)=(|{\bm k}|\cos \varphi,|{\bm k}|\sin \varphi$).
In the second equation of Eq.~(\ref{eq:heffkf}), we have approximated $|{\bm k}|$ with the Fermi wave number $k_{\rm F}$ assuming that the spin-orbit interaction energies, $k_{\rm F}\alpha$ and $k_{\rm F}\beta$, are much smaller than the Fermi energy\footnote{In semiconductor heterostructures, $k_{\rm F}\alpha$ and $k_{\rm F}\beta$ are typically less than $1\,{\rm meV}$.
For example, $k_{\rm F}\alpha$ was estimated as $0.07\,{\rm meV}$ in GaAs/AlGaAs heterostructures~\cite{Miller2003} using the electron density $n=5.0\times 10^{15}\, {\rm m}^{-2}$ and the effective mass $m^{*} = 0.067m_{\rm e}$\cite{Masselink1985}.
In addition, we obtain the Fermi wave number $k_{\rm F}\simeq 1.8\times 10^8\,{\rm m}^{-1}$ and the Fermi energy $\epsilon_{\rm F}\simeq 20\,{\rm meV}$ using the same electron density. We note that $\epsilon_{\rm F}\gg k_{\rm F}\alpha$ holds well.}.
When only the Rashba spin-orbit interaction exists 
($\beta = 0$), the energy band is spin-splitted as shown in Fig.~\ref{fig:band}~(a).
The spin polarization of each band depends on the azimuth angle $\varphi$ because it is determined by the effective Zeeman field ${\bm h}_{\rm eff}$ which is a function of $\varphi$ as seen in Eq.~(\ref{eq:heffkf}).
In the special case of $\beta/\alpha=1$, the spin polarization always becomes parallel to the direction of the azimuth angle $3\pi/4$ in the $xy$ plane as shown in Fig.~\ref{fig:band}~(b).
Then, the spin component in this direction is conserved.
This observation indicates that effect of 
the spin conservation may become important when the two spin-orbit interactions compete ($\alpha \simeq \beta$).

The Hamiltonian of the impurity potential is given as
\begin{align}
H_{\rm imp} &= u \sum_{i\in {\rm imp}} \sum_{\sigma}{\Psi_{\sigma}^\dagger({\bm r}_i) \Psi_\sigma({\bm r}_i)}, 
\end{align}
where $\Psi_{\sigma}({\bm r}) = {\cal A}^{-1/2} \sum_{{\bm k}} c_{{\bm k}\sigma} e^{i {\bm k}\cdot {\bm r}}$, ${\cal A}$ is the area of the junction, $u$ is the strength of the impurity potential, and ${\bm r}_i$ is the position of the impurity site.

The finite-temperature Green's function for the conduction electrons is defined by a $2\times 2$ matrix $\hat{g}({\bm k},i\omega_m)$ whose elements are
\begin{align}
g_{\sigma \sigma'}({\bm k},i\omega_m) &= \int_0^{\hbar \tempinv} d\tau \, e^{i\omega_m \tau} g_{\sigma \sigma'}({\bm k},\tau), \\
g_{\sigma \sigma'}({\bm k},\tau) &=
-\hbar^{-1} \langle c_{{\bm k}\sigma}(\tau) c_{{\bm k}\sigma'}^\dagger \rangle, 
\end{align}
where $c_{{\bm k}\sigma}(\tau) = e^{H_{\rm NM}\tau/\hbar} c_{{\bm k}\sigma} e^{-H_{\rm NM}\tau/\hbar}$, $H_{\rm NM} = H_{\rm kin}+H_{\rm imp}$, $\omega_m=\pi (2m+1)/\hbar\tempinv$ is the fermionic Matsubara frequency, and $\tempinv$ is the inverse temperature.
By employing the Born approximation, the finite-temperature Green's function can be expressed as
\begin{align}
& \hat{g}({\bm k},i\omega_m) = \frac{(i \hbar\omega_m  -\xi_{\bm k}+i \Gamma {\rm sgn}(\omega_m)/2)\hat{I} - {\bm h}_{\rm eff}\cdot {\bm \sigma}}{\prod_{\nu=\pm} (i\hbar\omega_m-E_{\bm k}^\nu +i\Gamma {\rm sgn}(\omega_m)/2)},
\label{eq:gkwimp}
\end{align}
where $E_{\bm k}^{\pm} = \xi_{\bm k} \pm |{\bm h}_{\rm eff}(\varphi)|$ is the spin-dependent electron dispersion, 
\begin{align}
\Gamma = 2\pi n_i u^2 D(\epsilon_{\rm F}) \label{def:Gamma}
\end{align}
is level broadening, and $n_i$ is the impurity concentration (see Appendix \ref{app:green} and Ref.~\onlinecite{Yama2021} for detailed derivation).

As already mentioned, the case of $\beta/\alpha=1$ is special because the spin component parallel to the direction of the azimuth angle $3\pi/4$ in the $xy$ plane is conserved (see Fig.~\ref{fig:band}~(b)).
By defining the spin component in this direction as 
\begin{align}
s^{3\pi/4}_{\rm tot} &\equiv \frac{1}{2}\sum_{\bm{k}}(c^{\dagger}_{\bm{k}+}c_{\bm{k}+}-c^{\dagger}_{\bm{k}-}c_{\bm{k}-}), \\
\left( \begin{array}{c} c_{\bm{k}+} \\ c_{\bm{k}-} \end{array} \right) &= \left( \begin{array}{cc} 1/\sqrt{2} & e^{-i3\pi/4}/\sqrt{2} \\ -e^{i3\pi/4}/\sqrt{2} & 1/\sqrt{2} \end{array} \right) \left( \begin{array}{c} c_{\bm{k}\uparrow} \\ c_{\bm{k}\downarrow} \end{array} \right),
\end{align}
we can prove $[H_{\rm kin} + H_{\rm imp}, s^{3\pi/4}_{\rm tot}] = 0$.
When the value of $\beta/\alpha$ is slightly shifted from 1, the spin conservation law is broken slightly and this leads to a slow spin relaxation. 
As will be discussed in Secs.~\ref{sec:effGIL} and \ref{sec:shift}, this slow spin relaxation, which is a remnant of the spin conservation at $\beta/\alpha = 1$, strongly affects the spin injection from the FI into the 2DEG.
To describe this feature, we need to consider the vertex correction to take the conservation law into account in our calculation as explained in Sec.~\ref{sec:VC}.

\subsection{Ferromagnetic insulator}

We consider the quantum Heisenberg model for the FI and employ the spin-wave approximation assuming that the temperature is much lower than the magnetic transition temperature and the magnitude of the localized spins, $S_0$, is sufficiently large. We write the expectation value of the localized spins in the FI as $\langle \bm{S}\rangle$, whose direction is $(\cos\theta,\sin\theta,0)$ as shown in the Fig.~\ref{fig:setup}~(b).
Using the Holstein-Primakov transformation, the Hamiltonian in the spin-wave approximation is obtained as
\begin{align}
    H_{{\rm FI}} &= \sum_{\bm k} \hbar \omega_{\bm k} b_{\bm k}^\dagger b_{\bm k}, 
    \label{eq:HFI1}
\end{align}
where $b_{\bm k}$ is the magnon annihilation operator with wave number ${\bm k}$, $\hbar \omega_{\bm k} = {\cal D}{\bm k}^2 + \hbar \gamma h_{\rm dc}$ is the energy dispersion of a magnon, ${\cal D}$ is the spin stiffness, $\gamma$ is the gyromagnetic ratio, and $h_{\rm dc}$ is the externally applied DC magnetic field.
We note that the external DC magnetic field controls the direction of the ordered spins.
We introduce new coordinates $(x',y',z')$ fixed on the ordered spins by rotating the original coordinates $(x,y,z)$ as shown in Fig.~\ref{fig:setup}~(b).
Then, the magnon annihilation operator is related to the spin ladder operator by the Holstein-Primakov transformation as $S^{x'+}_{\bm k}\equiv S^{y'}_{\bm{k}}+iS^{z'}_{\bm{k}} = (2S_0)^{1/2} b_{\bm k}$. 
The spin correlation function is defined as
\begin{align}
G_0({\bm k},i\omega_n) 
&= \int_0^{\hbar \tempinv} d\tau \, e^{i\omega_n \tau} G_0({\bm k},\tau),\\
G_0({\bm k},\tau) 
&= -\frac{1}{\hbar} \langle 
    S^{x'+}_{{\bm k}}(\tau) S^{x'-}_{{\bm k}}(0)
    \rangle ,
\end{align}
where $\omega_n = 2n\pi/\hbar\tempinv$ is the bosonic Matsubara frequency. 
The spin correlation function is calculated from the Hamiltonian (\ref{eq:HFI1}), as
\begin{align}
G_0({\bm k},i\omega_n) & =\frac{2S_0/\hbar}{i\omega_n-\omega_{\bm k}-\alpha_{\rm G}|\omega_n|}, \label{eq:correlation_magnon}
\end{align}
where $\alpha_{\rm G}>0$ is a phenomenological dimensionless parameter that describes the strength of the Gilbert damping in the bulk FI.

\subsection{Effect of the FI/2DEG interface}

The coupling between the FI and 2DEG can be accounted for by the Hamiltonian,
\begin{align}
H_{\rm int} &= \sum_{{\bm k}} ({\cal T}_{\bm k} S_{\bm k}^{x'+} s_{\bm k}^{x'-} + {\cal T}_{\bm k}^* s_{\bm k}^{x'+} S_{\bm k}^{x'-}),
\end{align}
where ${\cal T}_{\bm k}$ is an exchange interaction at a clean interface, for which the momentum of spin excitation is conserved.
The spin ladder operators for conduction electrons, $s_{\bm k}^{x'\pm}$, are obtained using a coordinate rotation as~\cite{Yama2021}
\begin{align}
s_{\bm k}^{x'\pm} &=
\frac12 \sum_{\sigma,\sigma'} \sum_{{\bm k}'} c_{{\bm k}'\sigma}^\dagger
(\hat{\sigma}^{x'\pm})_{\sigma\sigma'} c_{{\bm k}'\pm {\bm k}\sigma'},
\label{eq:skpm} \\
\hat{\sigma}^{x'\pm} &= -\sin \theta \, \sigma_x + \cos \theta \, \sigma_y \pm i \sigma_z ,
\label{eq:skmmatrix}
\end{align}
where $\hat{\sigma}^{x'\pm}\equiv \hat{\sigma}_{y'}\pm i\hat{\sigma}_{z'}$ and
\begin{equation*}
\begin{pmatrix} \hat{\sigma}_{x'} \\ \hat{\sigma}_{y'} \\ \hat{\sigma}_{z'}
\end{pmatrix}
=
\begin{pmatrix} 
\cos\theta & \sin\theta & 0
\\ -\sin\theta & \cos\theta & 0\\
0 & 0 & 1
\end{pmatrix}
\begin{pmatrix} \sigma_{x} \\ \sigma_{y} \\ \sigma_{z}
\end{pmatrix}.
\end{equation*}
Assuming that the interfacial exchange interaction is much smaller than the spin-orbit interactions, $k_{\rm F}\alpha$ and $k_{\rm F}\beta$\cite{NOGUES1999203,footnoteapp}, we perform a second-order perturbation theory with respect to the interfacial exchange interaction $H_{\rm int}$.
Accordingly, the spin correlation function of the FI is calculated as
\begin{align}
G({\bm k},i\omega_{n}) &= \frac{1}{(G_0({\bm k},i\omega_n))^{-1}-\Sigma({\bm k},i\omega_n)}, \\
\Sigma({\bm k},i\omega_n) 
&=|\mathcal{T}_{\bm k}|^2 {\cal A} \chi({\bm k},i\omega_n), \label{eq:SelfEnergyChi}
\end{align}
where $\Sigma({\bm k},i\omega_n)$ is the self-energy due to the interfacial exchange coupling and $\chi({\bm k},i\omega_n)$ is the spin susceptibility for conduction electrons per unit area, defined as
\begin{align}
\chi({\bm k},i\omega_n) &= \int_0^{\hbar \tempinv} d\tau \, e^{i\omega_n \tau} \chi({\bm k},\tau), \\
\chi({\bm k},\tau) &= - \frac{1}{\hbar {\cal A}} \langle s_{\bm k}^{x'+}(\tau) s_{\bm k}^{x'-}(0) \rangle,
\end{align}
where $s_{\bm k}^{x'\pm}(\tau)=e^{H_{\rm NM}\tau/\hbar}s_{\bm k}^{x'\pm} e^{-H_{\rm NM}\tau/\hbar}$.
Within the second-order perturbation, we only need to calculate the spin susceptibility for pure 2DEG without considering the junction because the interfacial coupling is already taken into account in the prefactor of the self-energy in Eq.~(\ref{eq:SelfEnergyChi}).
The uniform component of the retarded spin correlation function is obtained by analytic continuation $i\omega_n \rightarrow \omega + i\delta$, as
\begin{align}
G^R({\bm 0},\omega) 
&= \frac{2S_{0}/\hbar}{\omega-(\omega_0+\delta \omega_0) + i(\alpha_{\rm G}+\delta \alpha_{\rm G}) \omega},\label{eq:Green's_magnon} \\
\frac{\delta \omega_0}{\omega_0} &\simeq \frac{2S_0|\mathcal{T}_{\bm 0}|^2 {\cal A} }{\hbar\omega_0}\, {\rm Re}\, \chi^R({\bm 0},\omega_0),
\label{eq:deltaomega} \\
\delta \alpha_{\rm G} &\simeq -\frac{2S_{0}|\mathcal{T}_{\bm 0}|^2 {\cal A}}{\hbar\omega_0}\, {\rm Im}\, \chi^R({\bm 0},\omega_0),
\label{eq:deltagamma}
\end{align}
where the superscript $R$ indicates the retarded component, $\omega_0 = \omega_{\bm k=\bm 0}$ ($=\gamma h_{\rm dc}$) is the FMR frequency, and 
$\delta \omega_0$ and $\delta\alpha_{\rm G}$ are respectively the changes in the FMR frequency and Gilbert damping due to the FI/2DEG interface. 
We note that in contrast with the bulk Gilbert damping $\alpha_{\rm G}$, the increase of the Gilbert damping, $\delta \alpha_{\rm G}$, can be related directly to the spin susceptibility of 2DEG as shown by Eq.~(\ref{eq:deltagamma}).
In fact, measurement of $\delta \alpha_{\rm G}$ has been utilized as a qualitative indicator of spin current through a junction~\cite{Wang2014,Yang2018}.
In Eqs.~(\ref{eq:deltaomega}) and (\ref{eq:deltagamma}), we made an approximation by replacing $\omega$ with the FMR frequency $\omega_0$ by assuming that the FMR peak is sufficiently sharp ($\alpha_{\rm G} + \delta \alpha_{\rm G} \ll 1$).
Thus, both the FMR frequency shift and the modulation of the Gilbert damping are determined by the uniform spin susceptibility of the conduction electrons, $\chi({\bm 0},\omega)$.
In what follows, we include the vertex correction for calculation of $\chi({\bm 0},\omega)$, which was not taken into account in our previous work~\cite{Yama2021}.

\section{Vertex Correction}
\label{sec:VC}

\begin{figure}[tb]
\centering
\includegraphics[width=70mm]{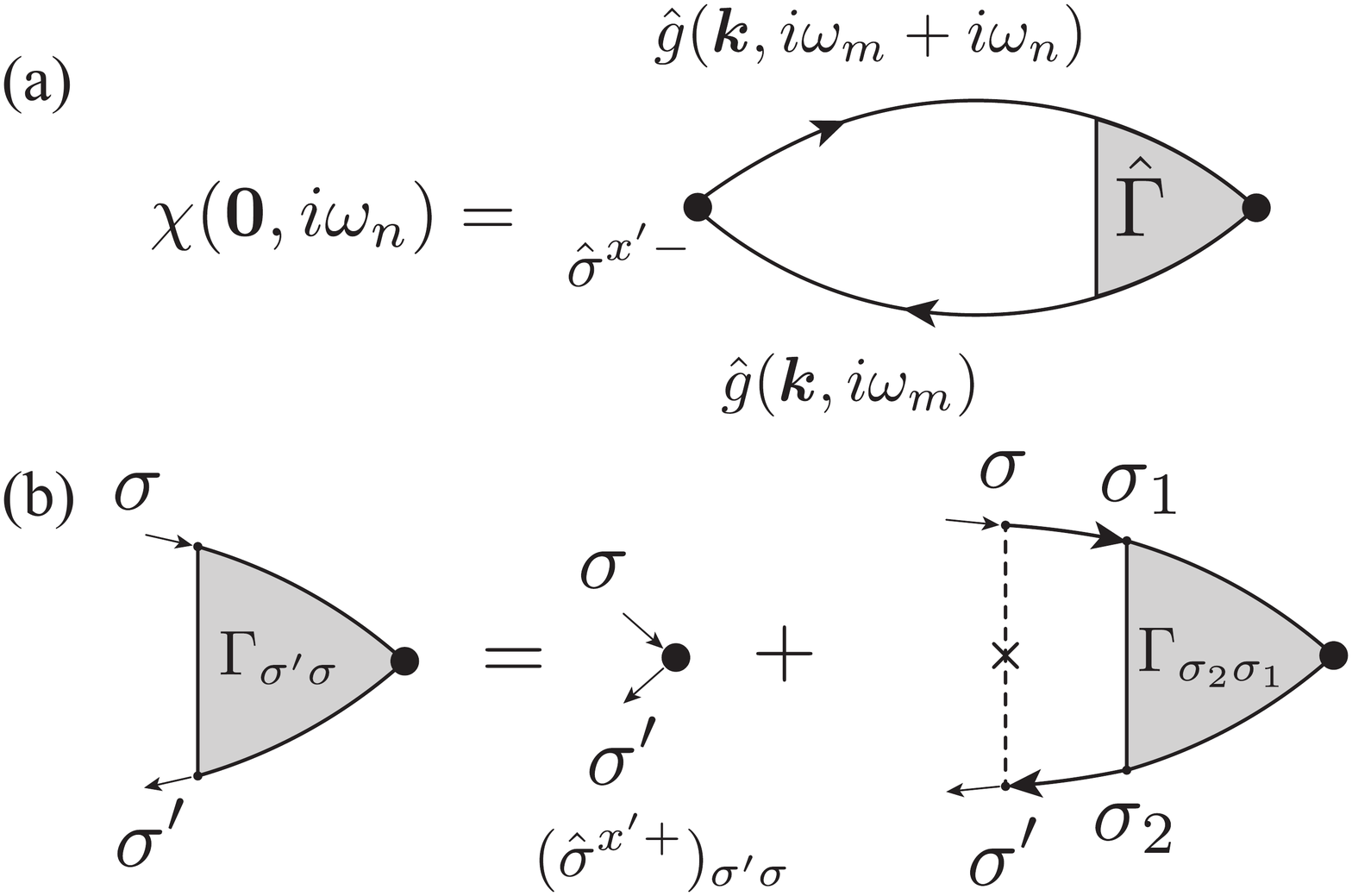}
\caption{Feynman diagrams of (a) the uniform spin susceptibility and (b) the Bethe-Salpeter equation for the ladder-type vertex function derived from the Born approximation.
The cross with two dashed lines indicates interaction between an electron and an impurity.}
\label{fig:BetheSalpeter}
\end{figure}

We calculate the spin susceptibility in the ladder approximation\cite{Bruus2004,Akkermans2007} that obeys the Ward-Takahashi relation with the self-energy in the Born approximation~\cite{Fulde1968}.
The Feynman diagrams for the corresponding spin susceptibility and the Bethe-Salpeter equation for the vertex function are shown in Figs.~\ref{fig:BetheSalpeter}~(a) and \ref{fig:BetheSalpeter}~(b), respectively.
The spin susceptibility of 2DEG is written as
\begin{align}
\chi({\bm 0},i\omega_n) &= \frac{1}{4\tempinv{\cal A}} 
\sum_{{\bm k},i\omega_m} {\rm Tr} \Bigl{[} 
\hat{g}({\bm k},i\omega_m)\hat{\Gamma}(\bm k,i\omega_{m},i\omega_{n}) \nonumber \\
&\hspace{12mm} \hat{g}({\bm k},i\omega_m+i\omega_n) \hat{\sigma}^{x'-} \Bigl{]},
\label{eq:susceptibility}
\end{align}
where the vertex function $\hat{\Gamma}(\bm k,i\omega_{m},i\omega_{n})$ is a $2\times 2$ matrix whose components are determined by the Bethe-Salpeter equation [see Fig.~\ref{fig:BetheSalpeter}~(b)],
\begin{align}
&{\Gamma}_{\sigma'\sigma}(\bm{k},i\omega_{m},i\omega_{n})  \nonumber \\
&= (\hat{\sigma}^{x'+})_{\sigma'\sigma} +\frac{u^{2}n_{i}}{{\cal A}}\sum_{\bm{q}}\sum_{\sigma_{1}\sigma_{2}}{g}_{\sigma'\sigma_{2}}(\bm{q},i\omega_{m})\nonumber \\
& \hspace{10mm} \times {\Gamma}_{\sigma_{2}\sigma_{1}}(\bm{q},i\omega_m,i\omega_n) 
{g}_{\sigma_{1} \sigma}(\bm{q},i\omega_{m}+i\omega_{n}).
\label{eq:BEtmp}
\end{align}
Since the right-hand side of this equation is independent of ${\bm k}$, the vertex function can simply be described as $\hat{\Gamma}(i\omega_m,i\omega_n)$.
We express the vertex function with the Pauli matrices as
\begin{align}
\hat{\Gamma}(i\omega_{m},i\omega_{n}) &\equiv  E\hat{I} + X\hat{\sigma}_{x'}+Y\hat{\sigma}_{y'}+Z\hat{\sigma}_{z'}, \label{eq:GammaE'X'Y'Z'}
\end{align}
where $E$, $X$, $Y$, and $Z$ will be determined self-consistently later.
The Green's function for the conduction electrons can be rewritten as
\begin{align}
\hat{g}(\bm{q},i\omega_{m})&=
\frac{A\hat{I}+B\hat{\sigma}_{x'}+C\hat{\sigma}_{y'}}{D}, \label{eq:ghatsigma} \\
A(i\omega_m) &= i\hbar\omega_{m}-\xi_{\bm{q}}+\frac{i\Gamma}{2}{\rm sgn}(\omega_{m}),\\
B &= -h_{\rm eff} \cos (\phi-\theta), \\
C &= -h_{\rm eff} \sin (\phi-\theta), \\
D(i\omega_m) &= \prod_{\nu=\pm} [i\hbar\omega_{m}-E^\nu_{\bm{q}}+\frac{i\Gamma}{2}{\rm sgn}(\omega_{m})] ,
\end{align}
where $\phi$ is the azimuth angle by which the effective Zeeman field is written as $\bm{h}_{\rm eff}=(h_{\rm eff}\cos\phi,h_{\rm eff}\sin\phi,0)$. This $h_{\rm eff}$ is written as $h_{\rm eff}
\simeq k_{\rm F}\sqrt{\alpha^2+\beta^2+2\alpha\beta\sin2\varphi}$ using the Fermi wave number $k_{\rm F}$. By substituting Eqs.~(\ref{eq:GammaE'X'Y'Z'}) and (\ref{eq:ghatsigma}) into the second term of Eq.~(\ref{eq:BEtmp}) and by the algebra of Pauli matrices, we obtain
\begin{align}
&\frac{u^{2}n_{i}}{{\cal A}}\sum_{\bm{q}} \hat{g}(\bm{q},i\omega_{m}) \hat{\Gamma}(\bm{q},i\omega_m,i\omega_n) 
\hat{g}(\bm{q},i\omega_{m}+i\omega_{n}) \nonumber \\
& = E' \hat{I} + X' \hat{\sigma}_{x'} + Y' \hat{\sigma}_{y'} + Z' \hat{\sigma}_{z'},
\label{eq:BEtmp2}
\end{align}
where
\begin{align}
\begin{pmatrix}
E' \\ X' \\ Y' \\ Z'
\end{pmatrix}
= \begin{pmatrix}
\Lambda_0 + \Lambda_1 & 0 & 0 & 0 \\
0 & \Lambda_0 + \Lambda_2 & \Lambda_3 & 0 \\
0 & \Lambda_3 & \Lambda_0-\Lambda_2 & 0 \\
0 & 0 & 0 & \Lambda_0 - \Lambda_1
\end{pmatrix}
\begin{pmatrix}
E \\ X \\ Y \\ Z
\end{pmatrix} ,
\label{primetononprime}
\end{align}
and $\Lambda_j(i\omega_m,i\omega_n)$ ($j=0,1,2,3$) are expressed as
\begin{align}
\Lambda_0(i\omega_m,i\omega_n) &= \frac{u^2 n_i}{\cal A} \sum_{\bm q} \frac{AA'}{DD'}, 
\label{Lambda0} \\
\Lambda_1(i\omega_m,i\omega_n) &=  \frac{u^2 n_i}{\cal A} \sum_{\bm q} \frac{h_{\rm eff}^2}{DD'}, \\
\Lambda_2(i\omega_m,i\omega_n) &=\frac{u^2 n_i}{\cal A} \sum_{\bm q} \frac{h_{\rm eff}^2\cos 2(\phi-\theta)}{DD'}, \\
\Lambda_3(i\omega_m,i\omega_n) &=\frac{u^2 n_i}{\cal A} \sum_{\bm q} \frac{h_{\rm eff}^2\sin 2(\phi-\theta)}{DD'},
\label{Lambda3}
\end{align}
using the abbreviated symbols, $A=A(i\omega_{m})$, $A'=A(i\omega_{m}+i\omega_{n})$, $D=D(i\omega_{m})$, and
$D'= D(i\omega_{m}+i\omega_{n})$.
Here, we have used the fact that the contributions of the first-order terms of $B$ and $C$ become zero after replacing the sum with the integral with respect to $\bm q$ and performing the azimuth integration. 
We can solve for $E$, $X$, $Y$, and $Z$ by combining Eq.~(\ref{primetononprime}) and the Bethe-Salpeter equation (\ref{eq:BEtmp}), which we rewrite as
\begin{align}
&  E \hat{I} + X \hat{\sigma}_{x'} + Y \hat{\sigma}_{y'} + Z \hat{\sigma}_{z'} \nonumber \\
& = \hat{\sigma}^{x'+}+E' \hat{I} + X' \hat{\sigma}_{x'} + Y' \hat{\sigma}_{y'} + Z' \hat{\sigma}_{z'},
\end{align}
with $\hat{\sigma}^{x'+} = \hat{\sigma}_{y'} + i \hat{\sigma}_{z'}$.
The solution is 
\begin{align}
E &= 0,\\
X &= \frac{\Lambda_3}{(1-\Lambda_0)^2-\Lambda_2^2{-\Lambda_3^2}}, 
\label{eq:X} \\
Y &= {\frac{1-\Lambda_0-\Lambda_2}{(1-\Lambda_0)^2-\Lambda_2^2-\Lambda_3^2}}, \\
Z &= \frac{i}{1-\Lambda_0+\Lambda_1}.
\label{eq:Z}
\end{align}
By replacing the sum with an integral as $\xi\equiv \xi_{\bm{q}}$,
\begin{align}
\frac{1}{\mathcal{A}}\sum_{\bm q}(\cdots)\simeq
D(\epsilon_{\rm F})\int_{-\infty}^{\infty}d\xi\int_{0}^{2\pi}\frac{d\varphi}{2\pi}(\cdots),
\label{eq:ksumformula}
\end{align}
Eqs.~(\ref{Lambda0})-(\ref{Lambda3}) can be rewritten as
\begin{align}
\Lambda_j(i\omega_m,i\omega_n) &= \theta(-\omega_{m})\theta(\omega_{m}+\omega_{n}) \tilde{\Lambda}_j(i\omega_n), 
\label{eq:Lambdaproperty}
\\
\tilde{\Lambda}_j(i\omega_n) &= \frac{i\Gamma}{4}\int_{0}^{2\pi}\frac{d\varphi}{2\pi}\nonumber \\
& \times \sum_{\nu,\nu'=\pm}\frac{f_j(\nu,\nu',\varphi)}{i\hbar\omega_{n}+(\nu-\nu')h_{\rm eff}(\varphi)+i\Gamma},
\label{eq:Lambdajtildedef}
\end{align}
where we have used Eq.~(\ref{def:Gamma}),
$\theta(x)$ is a step function, and
\begin{align}
f_0(\nu,\nu',\varphi) &= 1,\\
f_1(\nu,\nu',\varphi) &= \nu \nu',\\
f_2(\nu,\nu',\varphi) &= \nu \nu' \cos 2(\phi(\varphi)-\theta),\\
f_3(\nu,\nu',\varphi) &= \nu \nu' \sin 2(\phi(\varphi)-\theta).
\label{eq:f3def}
\end{align}
For detailed derivation, see Appendix~\ref{app:LambdaCalc}.
Substituting the Green's function and the vertex function into Eq.~(\ref{eq:susceptibility}), we obtain
\begin{align}
\chi({\bm 0},i\omega_n) 
&=\frac{1}{4\tempinv {\cal A}} 
\sum_{{\bm k},i\omega_m}\frac{2}{DD'}\Bigl{[} 2BCX
 \nonumber \\
& \hspace{-10mm} + (AA'-B^{2}+C^{2})Y -i(AA'-B^{2}-C^{2})Z
\Bigl{]}. \label{eq:chibfranaly}
\end{align}
By summing over ${\bm k}$ and $\omega_m$ and by analytical continuation, $i\omega_n \rightarrow \omega+i\delta$, the retarded spin susceptibility is obtained as\footnote{We note that the uniform spin susceptibility given in Eq.~(\ref{eq:chiend}) becomes independent of the temperature if the density of state of 2DEG is assumed to be constant. Although we can derive its temperature-dependent correction by the Sommerfeld expansion, it is small as long as $k_{\rm B}T$ is much smaller than the Fermi energy.}
\begin{align}
&\chi^R({\bm 0},\omega) \nonumber \\
&=\frac{D(\epsilon_{\rm F})\hbar\omega}{2i\Gamma}
\biggl{[}
\frac{\tilde{\Lambda}^{R}_0(1-\tilde{\Lambda}^{R}_0)-\tilde{\Lambda}^{R}_2(1-\tilde{\Lambda}^{R}_2)+(\tilde{\Lambda}^{R}_3)^{2}}{(1-\tilde{\Lambda}^{R}_0)^2-(\tilde{\Lambda}^{R}_2)^2{-(\tilde{\Lambda}^{R}_3)^2}}
\nonumber \\
&\hspace{10mm} +\frac{\tilde{\Lambda}_0^R-\tilde{\Lambda}^{R}_1}{1-\tilde{\Lambda}^{R}_0+\tilde{\Lambda}^{R}_1}
\biggl{]}-D(\epsilon_{\rm F}),\label{eq:chiend}
\end{align}
where 
\begin{align}
\tilde{\Lambda}^{R}_j &= \tilde{\Lambda}^R_j(\omega) = \tilde{\Lambda}_j(i\omega_n\rightarrow \omega+i\delta) \nonumber \\
&=\frac{i\Gamma}{4\Delta_0}\int_{0}^{2\pi}\frac{d\varphi}{2\pi} \nonumber \\
&\hspace{5mm} \times \sum_{\nu\nu'}\frac{f_j(\nu,\nu',\varphi)}{\hbar\omega/\Delta_0+(\nu-\nu')h_{\rm eff}/\Delta_0+i\Gamma/\Delta_0}.\label{eq:TildeuRdjfj}
\end{align}
A detailed derivation is given in Appendix~\ref{app:analytic}.
Here, we have introduced a unit of energy, $\Delta_0 = k_{\rm F}\alpha$, for the convenience of making the physical quantities dimensionless.
Using Eqs.~(\ref{eq:deltaomega}) and (\ref{eq:deltagamma}), we finally obtain the shift in the FMR frequency and the modulation of the Gilbert damping as
\begin{align}
\frac{\delta \omega_0}{\omega_0} &= \alpha_{{\rm G},0} \, {\rm Re} \, F(\omega_0), \label{eq:resltome}\\
\delta \alpha_{\rm G} &= -\alpha_{{\rm G},0} \, {\rm Im} \, F(\omega_0), \label{eq:resltgil}\\
F(\omega) &=\frac{\Delta_0}{2\pi i\Gamma}
\biggl{[}
\frac{\tilde{\Lambda}^{R}_0(1-\tilde{\Lambda}^{R}_0)-\tilde{\Lambda}^{R}_2(1-\tilde{\Lambda}^{R}_2)+(\tilde{\Lambda}^{R}_3)^{2}}{(1-\tilde{\Lambda}^{R}_0)^2-(\tilde{\Lambda}^{R}_2)^2{-(\tilde{\Lambda}^{R}_3)^2}}
\nonumber \\
&\hspace{10mm} +\frac{\tilde{\Lambda}_0^R-\tilde{\Lambda}^{R}_1}{1-\tilde{\Lambda}^{R}_0+\tilde{\Lambda}^{R}_1}
\biggl{]}-\frac{\Delta_0}{\pi \hbar \omega},
\label{Fomegadef}
\end{align}
where $\alpha_{{\rm G},0} = 2\pi S_0|{\cal T}_{\bm 0}|^2 {\cal A}D(\epsilon_{\rm F})/\Delta_0$ is a dimensionless parameter that describes the coupling strength at the interface.
This is our main result.

The spin susceptibility without the vertex correction can be obtained by taking the first-order term with respect to $\tilde{\Lambda}^{R}_j$:
\begin{align}
& \chi^{R}({\bm 0},\omega)
\simeq \frac{\hbar \omega D(\epsilon_{\rm F})}{2i\Gamma}
\bigl[ 2\tilde{\Lambda}^{R}_0-\tilde{\Lambda}^{R}_1 -\tilde{\Lambda}^{R}_2 \bigr]-D(\epsilon_{\rm F}) \nonumber \\
&=  \hbar \omega D(\epsilon_{\rm F}) \int\frac{d\varphi}{2\pi}
\Bigl[ \frac{1}{\hbar \omega + i\Gamma} \frac{1-\cos^2 (\phi(\varphi)-\theta)}{2}\nonumber \\
& \hspace{3mm} +\frac{1}{\hbar \omega - 2h_{\rm eff}(\varphi) + i\Gamma} \frac{1+\cos^2 (\phi(\varphi)-\theta)}{4} \nonumber \\
& \hspace{3mm} +\frac{1}{\hbar \omega +2h_{\rm eff}(\varphi) + i\Gamma} \frac{1+\cos^2 (\phi(\varphi)-\theta)}{4}\Bigr]-D(\epsilon_{\rm F}).
\end{align}
The imaginary part of $\chi^{R}({\bm 0},\omega)$ reproduces the result of Ref.~\onlinecite{Yama2021}.
Using this expression, the shift in the FMR frequency and the modulation of the Gilbert damping without the vertex correction are obtained as
\begin{align}
\frac{\delta \omega_0^{\rm nv}}{\omega_0} &= \alpha_{{\rm G},0} \, {\rm Re} \, F_{\rm nv}(\omega_0), \label{eq:resltome2}\\
\delta \alpha_{\rm G}^{\rm nv} &= -\alpha_{{\rm G},0} \, {\rm Im} \, F_{\rm nv}(\omega_0), \label{eq:resltgil2}\\
F_{\rm nv}(\omega) &= \frac{\Delta_0}{2\pi i\Gamma}
\biggl{[} 2\tilde{\Lambda}^{R}_0 - \tilde{\Lambda}^{R}_1 - \tilde{\Lambda}^{R}_2
\biggl{]}-\frac{\Delta_0}{\pi \hbar \omega},
\end{align}

\section{Modulation of the Gilbert damping}
\label{sec:effGIL}

\begin{figure*}[tb]
\centering
\includegraphics[width=170mm]{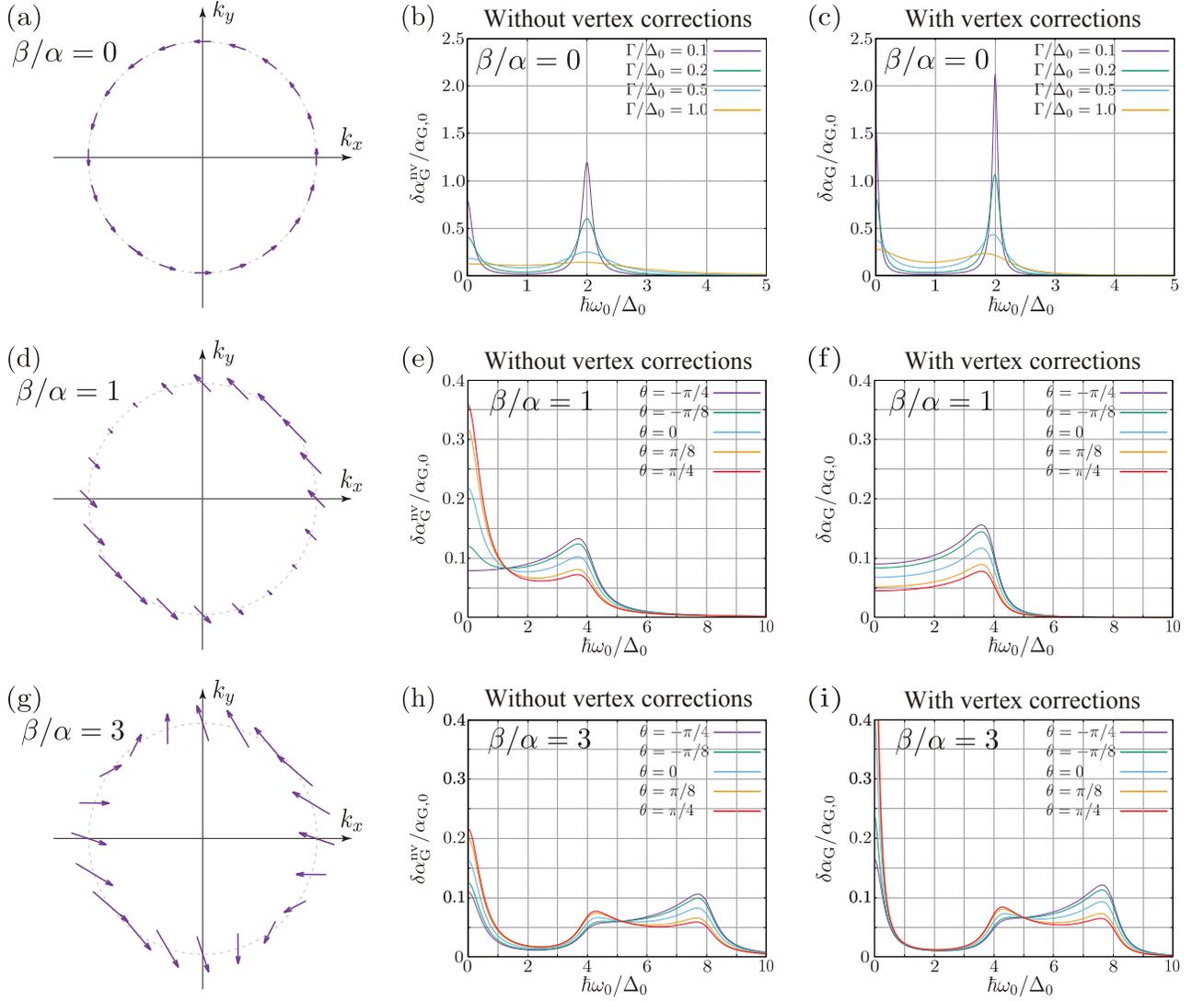}
\caption{(Left panels) Effective Zeeman field ${\bm h}_{\rm eff}$ on the Fermi surface. 
(Middle panels) Modulation of the Gilbert damping, $\delta \alpha^{\rm nv}_{\rm G}$, without vertex correction.
(Right panels) Modulation of the Gilbert damping with vertex correction, $\delta \alpha_{\rm G}$.
In the middle and right panels, the modulation of the Gilbert damping is plotted as a function of the FMR frequency, $\omega_0=\gamma h_{\rm dc}$.
The spin-orbit interactions are as follows. (a), (b), (c): $\beta/\alpha=0$. (d), (e), (f): $\beta/\alpha=1$. (g), (h), (i): $\beta/\alpha=3$.
We note that (b), (e), (h) are essentially the same result as Ref.~\onlinecite{Yama2021}.}
\label{fig:2Dplot}
\end{figure*}

First, we show the result for the modulation of the Gilbert damping, $\delta \alpha_{\rm G}$, for $\beta/\alpha = 0$, $1$, and $3$ and discuss the effect of the vertex correction by comparing it with the result without the vertex correction in Sec.~\ref{sec:vertexcorrectionresult1}.
Next, we discuss the strong enhancement of the Gilbert damping near $\beta/\alpha = 1$ in Sec.~\ref{sec:vertexcorrectionresult2}.

\begin{figure*}[tb]
\centering
\includegraphics[width=170mm]{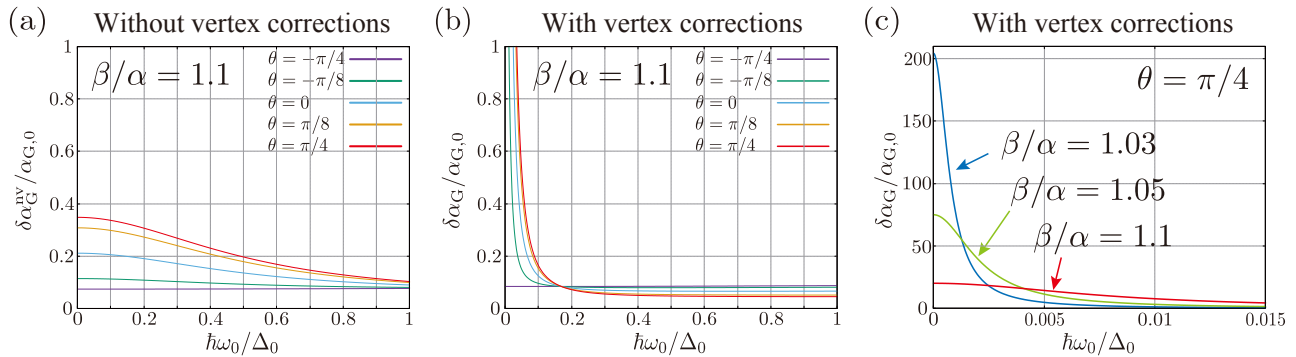}
\caption{Modulation of the Gilbert damping calculated for $\beta/\alpha=1.1$ (a) without the vertex correction and (b) with the vertex correction.
The horizontal axis is the FMR frequency $\omega_0$ and the five curves correspond to five different angles of $\langle {\bm S} \rangle$, i.e., $\theta = -\pi/4, -\pi/8, 0, \pi/8$, and $\pi/4$.
(c) Enlarged plot of the modulations of the Gilbert damping as a function of the FMR frequency $\omega_0$.
The angle of $\langle {\bm S} \rangle$ is fixed as $\theta=\pi/4$ and the three curves correspond to $\beta/\alpha=1.03$, $1.05$, and $1.1$.
In all the plots, we have chosen $\Gamma/\Delta_{0}=0.5$.}
\label{fig:GIL_Exact}
\end{figure*}

\subsection{Effect of vertex corrections}
\label{sec:vertexcorrectionresult1}

First, let us discuss the case of $\beta/\alpha=0$, i.e., the case when only the Rashba spin-orbit interaction exists\footnote{The result for the case of $\alpha/\beta = 0$, i.e., the case when only the Dresselhaus spin-orbit interaction exists, is the same as the case of $\beta/\alpha=0$}.
Figure~\ref{fig:2Dplot}~(a) shows the effective Zeeman field ${\bm h}_{\rm eff}$ along the Fermi surface.
Figures~\ref{fig:2Dplot}~(b) and \ref{fig:2Dplot}~(c) show the modulations of the Gilbert damping without and with the vertex correction.
The horizontal axes of Figs.~\ref{fig:2Dplot}~(b) and \ref{fig:2Dplot}~(c) denote the resonant frequency $\omega_0=\gamma h_{\rm dc}$
in the FMR experiment.
Note that the modulation of the Gilbert damping, $\delta \alpha_{\rm G}$, is independent of $\theta$, i.e., the azimuth angle of $\langle \bm S \rangle$.
The four curves in Figs.~\ref{fig:2Dplot}~(b) and \ref{fig:2Dplot}~(c) correspond to $\Gamma/\Delta_0 = 0.1$, $0.2$, $0.5$, and $1.0$\footnote{
The order of the electron mobility in GaAs/AlGaAs at low temperatures is $10^{5}~{\rm cm}^2/{\rm Vs}$\cite{Mendez1984} to $10^{7}~{\rm cm}^2/{\rm Vs}$\cite{Umansky1997}. By using this and the values in the footnote [58], we obtain that $\Gamma/\Delta_0$ is on the order of $10^{-2}$ to $1$.}.
We find that these two graphs have a common qualitative feature;
the modulation of the Gilbert damping has two peaks at $\omega_0=0$ and $\omega_0 = 2\Delta_0$ and their widths become larger as $\Gamma$ increases.
The peak at $\omega_0=0$ corresponds to elastic spin-flipping of conduction electrons induced by the transverse magnetic field via the exchange bias of the FI, while the peak at $\hbar\omega_0=2\Delta_0$ is induced by spin excitation of conduction electrons due to magnon absorption~\cite{Yama2021}.
In the case of $\beta/\alpha=0$, the vertex correction changes the modulation of the Gilbert damping moderately [compare Figs.~\ref{fig:2Dplot}~(c) with \ref{fig:2Dplot}~(b)].
The widths of the two peaks at $\omega_0=0$ and $\omega_0=2\Delta_0$ become narrower when the vertex correction is taken into account (see Appendix~\ref{app:alphazero} for the analytic expressions).

The case of $\beta/\alpha=1$ is special because the effective Zeeman field ${\bm h}_{\rm eff}$ always points in the direction of $(-1,1)$ or $(1,-1)$, as shown in Fig.~\ref{fig:2Dplot}~(d).
The amplitude of ${\bm h}_{\rm eff}$ depends on the angle of the wave number of the conduction electrons, $\varphi$, 
\begin{align}
h_{\rm eff}(\varphi)=2\Delta_0|\sin(\varphi+\pi/4)|,
\end{align}
and varies in the range of $0 \le 2h_{\rm eff} \le 4\Delta_0$.
Figures~\ref{fig:2Dplot}~(e) and \ref{fig:2Dplot}~(f) show the modulation of the Gilbert damping without and with the vertex correction for $\Gamma/\Delta_0 = 0.5$.
The five curves correspond to five different angles of $\langle {\bm S} \rangle$, $\theta = -\pi/4, -\pi/8, 0, \pi/8$, and $\pi/4$.
The most remarkable feature revealed by comparing Figs.~\ref{fig:2Dplot}~(f) with \ref{fig:2Dplot}~(e) is that the peak at $\omega_0=0$ disappears if the vertex correction is taken into account (see Appendix~\ref{app:alphaone} for the analytic expressions).
In the subsequent section, we will show that $\delta \alpha_{\rm G}(\omega_0)$ has a $\delta$-function-like singularity at $\omega_0 = 0$ for $\beta/\alpha=1$ due to the spin conservation law along the direction of ${\bm h}_{\rm eff}$.

In the case of $\beta/\alpha=3$, the direction of the effective Zeeman field ${\bm h}_{\rm eff}$ varies along the Fermi surface [Fig.~\ref{fig:2Dplot}~(g)].
Figures~\ref{fig:2Dplot}~(h) and \ref{fig:2Dplot}~(i) show the modulation of the Gilbert damping without and with the vertex correction for $\Gamma/\Delta_0 = 0.5$.
For $\beta/\alpha=3$, a peak at $\omega_0=0$ appears even when the vertex correction is taken into account.
The broad structure in the range of $4 \Delta_0 \le \hbar \omega_0 \le 8\Delta_0$ is caused by the magnon absorption process where its range reflects the distribution of the spin-splitting energy $2h_{\rm eff}$ along the Fermi surface.
By comparing Figs.~\ref{fig:2Dplot}~(h) and \ref{fig:2Dplot}~(i), we find that the vertex correction changes the result only moderately as in the case of $\beta/\alpha = 0$; 
the peak structure at $\omega_0 = 0$ becomes sharper when the vertex correction is taken into account while the broad structure is slightly enhanced.

\subsection{Strong enhancement of the Gilbert damping}
\label{sec:vertexcorrectionresult2}

Here, we examine the strong enhancement of the Gilbert damping for $\beta/\alpha\simeq 1$.
As explained in Sec.~\ref{sec:model}, the spin component in the direction of the azimuth angle $3\pi/4$ in the $xy$ plane is exactly conserved at $\beta/\alpha = 1$ [see also Fig.~\ref{fig:2Dplot}~(d)].
When the value of $\beta/\alpha$ is shifted slightly from 1, the spin conservation law is broken but the spin relaxation becomes remarkably slow.
To see this effect, we show the modulation of the Gilbert damping without and with the vertex correction for $\beta/\alpha = 1.1$ in Figs.~\ref{fig:GIL_Exact}~(a) and \ref{fig:GIL_Exact}~(b), respectively.
The five curves correspond to five different azimuth angles of $\langle {\bm S} \rangle$, and the energy broadening is set as $\Gamma/\Delta_0 = 0.5$.
Figs.~\ref{fig:GIL_Exact}~(a) and \ref{fig:GIL_Exact}~(b) indicate that the Gilbert damping is strongly enhanced at $\omega_0 = 0$ only when the vertex correction is taken into account.
This is the main result of our work.

\begin{figure}[htb]
\centering
\includegraphics[width=70mm]{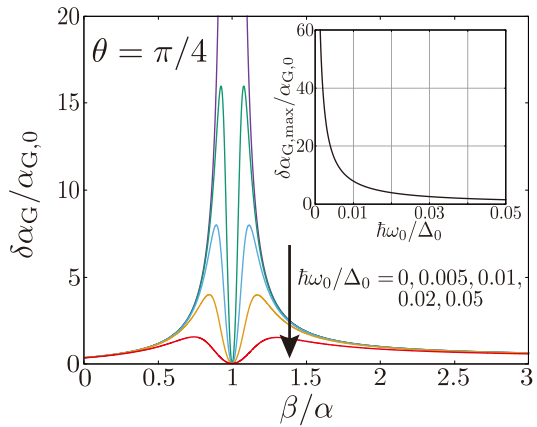}
\caption{Modulation of the Gilbert damping as a function of  $\beta/\alpha$.
The five curves correspond to $\hbar\omega_0/\Delta_0=0,~0.005$, $0.01$, $0.02$, and $0.05$.
We have taken the vertex correction into account and have chosen 
$\Gamma/\Delta_0=0.5$.
The inset illustrates maximum values of the modulation of the Gilbert damping, $\delta\alpha_{\rm G,max}$, in varying $\beta/\alpha$ for a fixed value of $\hbar\omega_0/\Delta_0$.
}
\label{fig:aog_deltaalpha}
\end{figure}

\begin{figure*}[tb]
\centering
\includegraphics[width=170mm]{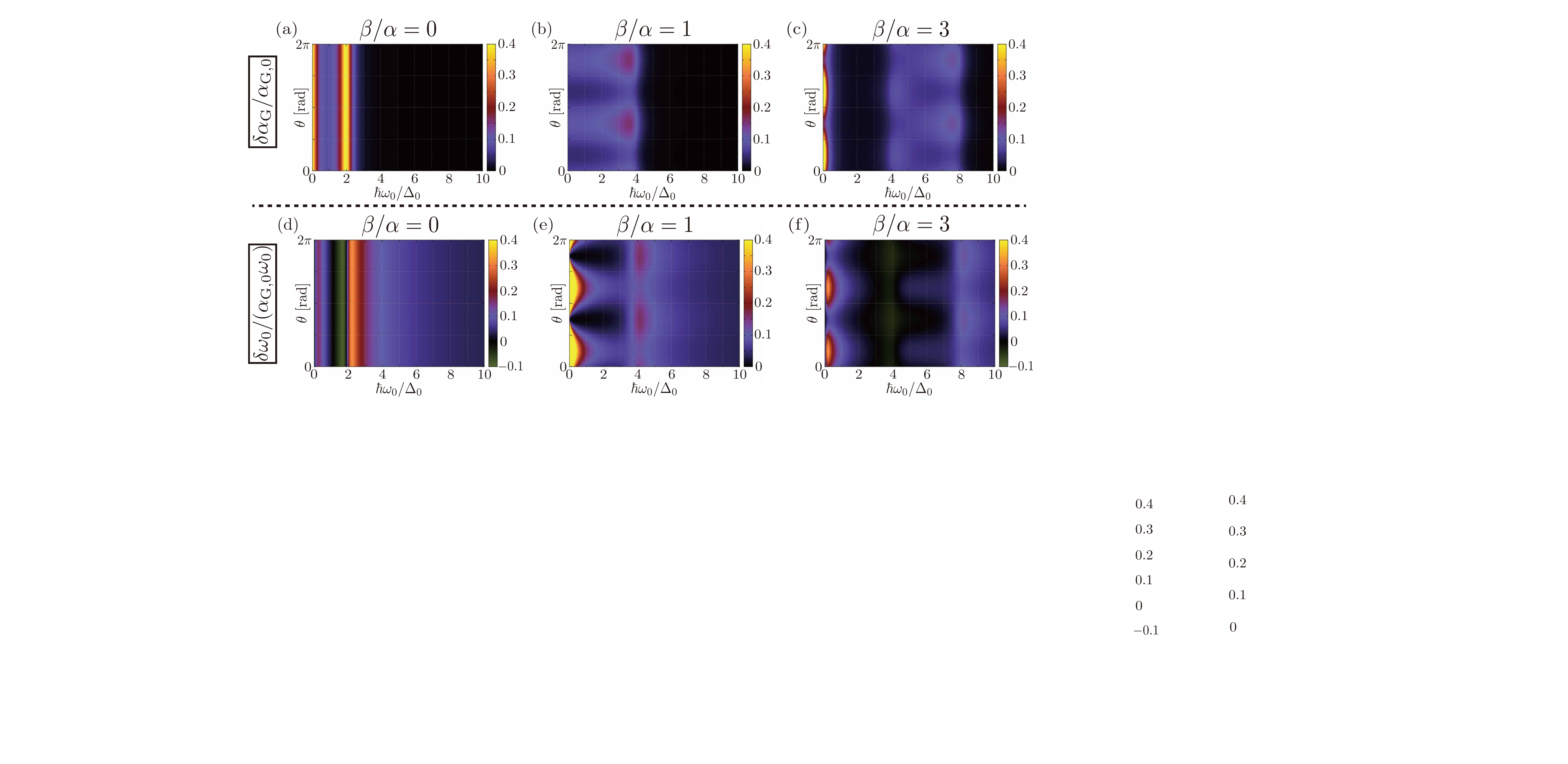}
\caption{(Upper panels) Modulations of the Gilbert damping, $\delta \alpha_{\rm G}/\alpha_{{\rm G},0}$ for (a) $\beta/\alpha=0$, (b) $\beta/\alpha=1$, and (c) $\beta/\alpha=3$.
(Lower panels) Shifts in the FMR frequency, $\delta \omega_0/(\alpha_{{\rm G},0}\omega_0)$, for (d) $\beta/\alpha=0$, (e) $\beta/\alpha=1$, and (f) $\beta/\alpha=3$.
The horizontal axes are the FMR frequency, $\omega_0=\gamma h_{\rm dc}$, while the vertical axes show the azimuth angle of the spontaneous spin polarization, $\theta$, in the FI. 
 In all the plots, we have considered vertex corrections and have chosen $\Gamma/\Delta_{0}=0.5$.
In (a), (c), and (e) there are regions in which the values exceed the upper limits of the color bar located in the right side of each plot; 
the maximum value is about $0.45$ in (a), $0.65$ in (c), and about $10$ in (e) (see also Fig.~\ref{fig:ENESHIFT}).
In addition, (b) cannot express a $\delta$-function-like singularity at $\omega_0 = 0$ (see the main text).}
\label{fig:density}
\end{figure*}

Figure~\ref{fig:GIL_Exact}~(c) plots the modulation of the Gilbert damping with the vertex correction for $\Gamma/\Delta_{0}=0.5$ and $\theta=\pi/4$, the latter of which corresponds to the case of the strongest enhancement at $\omega_0 = 0$.
The three curves correspond to $\beta/\alpha=1.03$, $1.05$, and $1.1$.
As the ratio of $\beta/\alpha$ approaches 1, the peak height at $\omega_0=0$ gets larger.
For $\beta/\alpha\simeq 1$, $\delta \alpha_{\rm G}$ is calculated approximately as
\begin{align}
\frac{\delta\alpha_{\rm G}}{\alpha_{{\rm G},0}}
&\simeq 
\frac{\Delta_{0}}{2\pi}
\frac{\Gamma_s}{(\hbar\omega_0)^{2}+\Gamma_s^{2}} \sin^{2}\Bigl{(} \theta+\frac{\pi}{4} \Bigl{)}, 
\label{eq:Approx1} \\
\Gamma_s &\equiv \frac{2}{\Gamma} \int_{0}^{2\pi}\frac{d\varphi}{2\pi}
\frac{(h_{x}+h_{y})^2}{1+(2h_{\rm eff}/\Gamma)^{2}}, 
\label{eq:Cend}
\end{align}
where $\Gamma_s$ gives the peak width in Figs.~\ref{fig:GIL_Exact}~(b) and \ref{fig:GIL_Exact}~(c) (see Appendix~\ref{app:nearone} for a detailed derivation).
For $\beta/\alpha=1+\delta$ ($\delta \ll 1$), $\Gamma_s$ is proportional to $\delta^2$ and approaches zero in the limit of $\delta \rightarrow 0$.
This indicates that $\Gamma_s$ corresponds to the spin relaxation rate due to a small breakdown of the spin conservation law away from the special point of $\beta/\alpha = 1$.
Note that the peak height of $\delta \alpha_{\rm G}$ at $\omega_0=0$ diverges at $\beta/\alpha = 1$.
This indicates that for $\beta/\alpha = 1$, $\delta \alpha_{\rm G}(\omega_0)$ has a $\delta$-function-like singularity at $\omega_0=0$, which is not drawn in Fig.~\ref{fig:2Dplot}~(f).

Figure~\ref{fig:aog_deltaalpha} plots the modulation of the Gilbert damping for $\Gamma/\Delta_{0}=0.5$ and $\theta=\pi/4$ as a function of $\beta/\alpha$.
The five curves correspond to $\hbar \omega_0/\Delta_0 = 0, 0.005, 0.01, 0.02$, and $0.05$, respectively.
This figure indicates that when we fix the resonant frequency $\omega_0$ and vary the ratio of $\beta/\alpha$, the Gilbert damping is strongly enhanced when $\beta/\alpha$ is slightly smaller or larger than 1.
We expect that this enhancement of the Gilbert damping is strong enough to be observed experimentally.
We note that $\delta\alpha_{\rm G}/\alpha_{\rm G,0}$ approaches $0.378$ ($0.318$) for $\beta/\alpha\rightarrow 0$ ($\beta/\alpha \rightarrow \infty$). The inset in Fig.~\ref{fig:aog_deltaalpha} plots maximum values of $\delta\alpha_{\rm G}/\alpha_{\rm G,0}$ when $\beta/\alpha$ is varied for a fixed value of $\hbar\omega_0/\Delta_0$.
In other words, the vertical axis of the inset corresponds to the peak height in the main panel for each value of $\hbar\omega_{0}/\Delta_0$.
We find that the maximum value of $\delta\alpha_{\rm G}/\alpha_{\rm G,0}$ diverges as $\omega_0$ approaches zero.

\begin{figure}[tb]
\centering
\includegraphics[width=75mm]{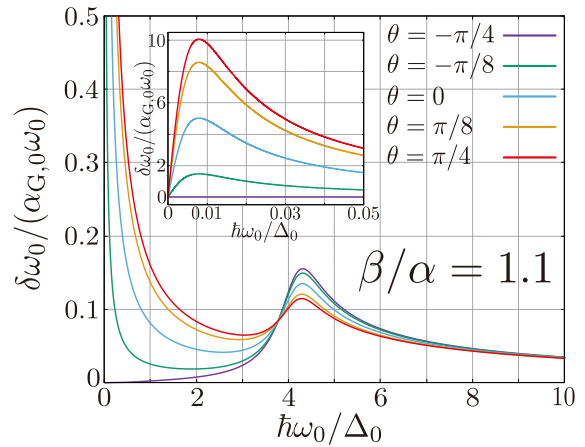}
\caption{Shift in FMR frequency, $\delta \omega_0/(\alpha_{{\rm G},0}\omega_0)$, as a function of the resonance frequency $\omega_0$ for $\beta/\alpha=1.1$. 
The inset shows the same quantities in the low-frequency range of $0 \le \hbar \omega_0/\Delta_0 \le 0.05$ with a larger scale on the vertical axis.
We have taken the vertex correction into account and have chosen $\Gamma/\Delta_0=0.5$.}
\label{fig:ENESHIFT}
\end{figure}

\section{Shift in the FMR frequency}
\label{sec:shift}

Next, we discuss the shift in the FMR frequency when the vertex correction is taken into account.
The density plots in Figs.~\ref{fig:density}~(a), \ref{fig:density}~(b), and \ref{fig:density}~(c) for $\beta/\alpha=0$, $1$, and $3$ summarize the modulation of the Gilbert damping, $\delta \alpha_{\rm G}$.
These plots have the same features as in Figs.~\ref{fig:2Dplot}~(c), \ref{fig:2Dplot}~(f), and \ref{fig:2Dplot}~(i).
Figures.~\ref{fig:density}~(d), \ref{fig:density}~(e), and \ref{fig:density}~(f) plot the shift in the FMR frequency $\delta \omega_0/\omega_0$ with density plots for $\beta/\alpha=0$, $1$, and $3$.
By comparing Figs.~\ref{fig:density}~(a), \ref{fig:density}~(b), and \ref{fig:density}~(c) with \ref{fig:density}~(d), \ref{fig:density}~(e), and \ref{fig:density}~(f), we find that some of the qualitative features of the FMR frequency shift are common to those of the modulation of the Gilbert damping, $\delta\alpha_{\rm G}$; (i) they depend on $\theta$ for $\beta/\alpha > 0$, while they do not depend on $\theta$ for $\beta/\alpha=0$, (ii) the structure at $\omega_0 = 0$ due to elastic spin-flipping appears, and (iii) the structure within a finite range of frequencies due to magnon absorption appears.
We can also see a few differences between $\delta \alpha_{\rm G}$ and $\delta \omega_0/\omega_0$.
For example, $\delta \omega_0/\omega_0$ has a dip-and-peak structure at $\hbar \omega_0/ \Delta_0 = 2$ where $\delta \alpha_{\rm G}$ has only a peak.
Related to this feature, $\delta \omega_0/\omega_0$ has a tail that decays more slowly than that for $\delta \alpha_{\rm G}$.
The most remarkable difference is that  $\delta \omega_0/\omega_0$ diverges at $\omega_0 = 0$ for $\beta/\alpha = 1$ except for $\theta = 3\pi/4,7\pi/4$, reflecting the $\delta$-function-like singularity of $\delta \alpha_{\rm G}$ at $\omega_0=0$.
These features are reasonable because $\delta \omega_0/\omega_0$ and $\delta \alpha_{\rm G}$, which are determined by the real and imaginary parts of the retarded spin susceptibility, are related to each other through the Kramers-Kronig conversion.

The main panel of Fig.~\ref{fig:ENESHIFT} shows the frequency shift $\delta\omega_0/\omega_0$ for $\beta/\alpha=1.1$ as a function of the resonant frequency $\omega_0$.
The five curves correspond to $\theta = -\pi/4, -\pi/8, 0, \pi/8$, and $\pi/4$.
Although the frequency shift appears to diverge in the limit of $\omega_0 \rightarrow 0$ in the scale of the main panel, it actually grows to a finite value and then goes to zero as $\omega_0$ approaches zero (see the inset of Fig.~\ref{fig:ENESHIFT}).
For $\beta/\alpha = 1 + \delta$ ($\delta \ll 1$), the frequency shift is calculated approximately as
\begin{align}
\frac{\delta \omega_0}{\alpha_{{\rm G},0} \omega_0}
\simeq \frac{\Delta_0}{2\pi}\frac{\hbar\omega_0}{(\hbar\omega_0)^{2}+\Gamma_s^{2}} \sin^{2}\Bigl{(}\theta+\frac{\pi}{4}\Bigl{)},
\label{eq:Approx2}
\end{align}
where $\Gamma_s$ is the spin relaxation rate defined in Eq.~(\ref{eq:Cend}) (see Appendix~\ref{app:nearone} for the detailed derivation).
We expect that this strong enhancement of the frequency shift near $\beta/\alpha = 1$ can be observed experimentally.

\section{Summary}
\label{sec:summary}

We theoretically investigated spin pumping into a two-dimensional electron gas (2DEG) with a textured effective Zeeman field caused by Rashba- and Dresselhaus-type spin-orbit interactions.
We expressed the change in the peak position and the linewidth in a ferromagnetic resonance (FMR) experiment that is induced by the 2DEG within a second-order perturbation with respect to the interfacial exchange coupling by taking the vertex correction into account.
The FMR frequency and linewidth are modulated by elastic spin-flipping or magnon absorption.
We found that, for almost all of the parameters, the vertex correction modifies the modulation of the Gilbert damping only moderately and does not change the qualitative features obtained in our previous paper~\cite{Yama2021}.
However, we found that the Gilbert damping at low frequencies, which is caused by elastic spin-flipping, is strongly enhanced when the Rashba- and Dresselhaus-type spin-orbit interactions are chosen to be almost equal but slightly different.
Even in this situation, the Gilbert damping at high frequencies, which is caused by magnon absorption, shows small modification.
This strong enhancement of the Gilbert damping at low frequencies appears only when the vertex correction is taken into account and is considered to originate from the slow spin relaxation related to the spin conservation law that holds when the two spin-orbit interactions completely match.
A similar enhancement was found for the frequency shift of the FMR due to elastic spin-flipping.
We expect that this remarkable enhancement can be observed experimentally.

Our work provides a theoretical foundation for spin pumping into two-dimensional electrons with a spin-textured Zeeman field on the Fermi surface.
Although we have treated a specific model for two-dimensional electron systems with both the Rashba and Dresselhaus spin-orbit interactions, our formulation and results will be helpful for describing spin pumping into general two-dimensional electron systems such as surface/interface states~\cite{Lesne2016,Diogo2019,Sanchez2013} and atomic layer compounds~\cite{Bangar2022,Dushenko2016}. 

\section*{Acknowledgements}
The authors thank Y. Suzuki, Y. Kato, and A. Shitade for helpful discussion. 
T. K. acknowledges support from the Japan Society for the Promotion of Science (JSPS KAKENHI Grant No.~JP20K03831). M. M. is financially supported by a Grant-in-Aid for Scientific Research B (Grants No.~JP20H01863, No.~JP21H04565, and No.~JP21H01800) from MEXT, Japan. M. Y. is supported by JST SPRING (Grant No. JPMJSP2108).

\appendix

\section{Calculation of Green's function}
\label{app:green}

\begin{figure}[tb]
\centering
\includegraphics[width=85mm]{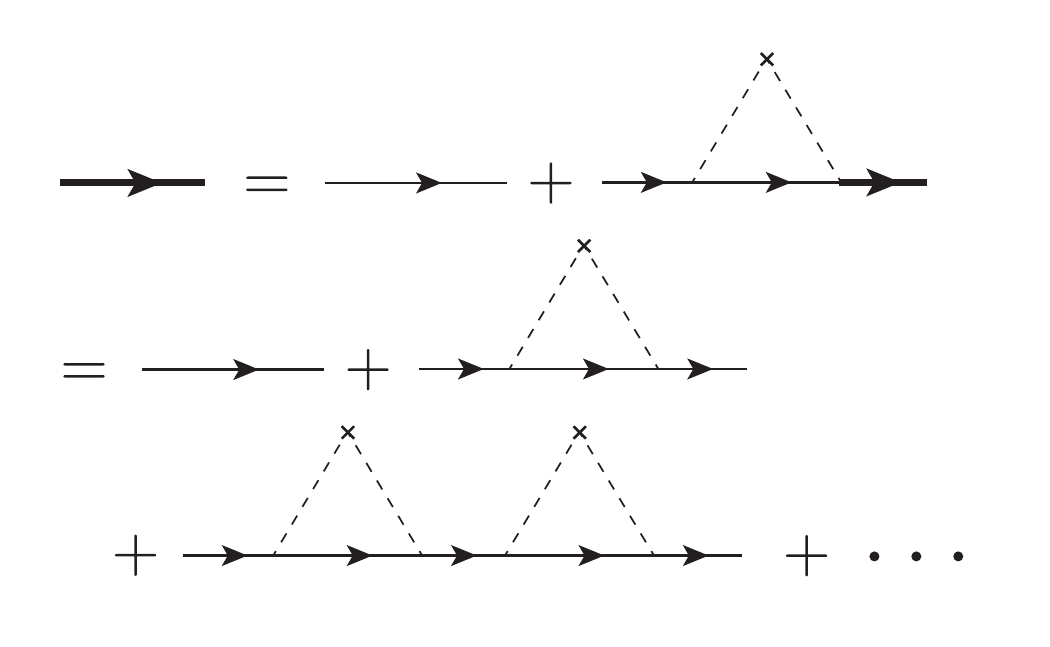}
\caption{The Feynman diagram for Green's function within the Born approximation.}
\label{fig:FeynmanG}
\end{figure}

In our work, Green's function of conduction electrons is calculated by taking effect of impurity scattering into account.
In general, the finite-temperature Green's function $\hat{g}({\bm k},i\omega_m)$ after the impurity average is described by the Dyson equation with the impurity self-energy $\hat{\Gamma}({\bm k},\omega_m)$ as
\begin{align}
\hat{g}({\bm k},i\omega_m) = \frac{1}{\hat{g}_0({\bm k},i\omega_m)^{-1}-\hat{\Gamma}({\bm k},i\omega_m)},
\end{align}
where $\hat{g}_0({\bm k},i\omega_m)^{-1}$ is Green's function of electrons in the absence of impurities.
In our work, we employ the Born approximation in which the self-energy is approximated by second-order perturbation with respect to an impurity potential.
In the Born approximation, the self-energy is given as
\begin{align}
\hat{\Gamma}({\bm k},i\omega_m) = n_i u^2 \int \frac{d^2{\bm k}}{(2\pi)^2} \hat{g}_0({\bm k},i\omega_m),
\end{align}
where $n_i$ is the impurity concentration.
The corresponding Feynman diagram of the Dyson equation is shown in Fig.~\ref{fig:FeynmanG}.
By straightforward calculation, Eq.~(\ref{eq:gkwimp}) can be derived.
For a detailed derivation, see Ref.~\onlinecite{Yama2021}.

\section{Derivation of Equations.~(\ref{eq:Lambdaproperty})-(\ref{eq:f3def})}
\label{app:LambdaCalc}

Eqs.~(\ref{Lambda0})-(\ref{Lambda3}) can be rewritten with $\Gamma = 2\pi n_i u^2 D(\epsilon_{\rm F})$ as
\begin{align}
\Lambda_0(i\omega_m,i\omega_n) &= \frac{i\Gamma}{4}
\int_0^{2\pi} \frac{d\varphi}{2\pi} \sum_{\nu,\nu'} I_{\nu\nu'}, \\
\Lambda_1(i\omega_m,i\omega_n) &= \frac{i\Gamma}{4}
\int_0^{2\pi} \frac{d\varphi}{2\pi} \sum_{\nu,\nu'} \nu \nu' I_{\nu\nu'}, \\
\Lambda_2(i\omega_m,i\omega_n) &= \frac{i\Gamma}{4}
\int_0^{2\pi} \frac{d\varphi}{2\pi} \cos2(\varphi-\theta) \sum_{\nu,\nu'} \nu \nu' I_{\nu\nu'}, \\
\Lambda_3(i\omega_m,i\omega_n) &= \frac{i\Gamma}{4}
\int_0^{2\pi} \frac{d\varphi}{2\pi} \sin 2(\varphi-\theta) \sum_{\nu,\nu'} \nu \nu' I_{\nu\nu'}, 
\end{align}
where 
\begin{align}
I_{\nu\nu'} &= \int_{-\infty}^\infty \frac{d\xi}{2 \pi i} 
\frac{1}{i\hbar \omega_m - \xi - \nu h_{\rm eff}+i(\Gamma/2){\rm sgn}(\omega_m)} \nonumber \\
&\times \frac{1}{i\hbar (\omega_m+\omega_n) - \xi - \nu' h_{\rm eff}+i(\Gamma/2){\rm sgn}(\omega_m+\omega_n)} .
\end{align}
We note that one needs to calculate this integral only for $\omega_n>0$ to obtain the retarded component by analytic continuation.
Then, we can easily prove by the residue integral that $I_{\nu\nu'}=0$ for $\omega_m > 0$ and $\omega_m + \omega_n > 0$ ($\omega_m < 0$ and $\omega_m + \omega_n < 0$) because both of the two poles in the integrand are located only in the upper (lower) half of the complex plane of $\xi$. 
For $\omega_m<0$ and $\omega_m+\omega_n>0$, the integral is evaluated by the residue integral as
\begin{align}
I_{\nu\nu'} &= \frac{1}{i\hbar \omega_n + (\nu-\nu')h_{\rm eff} + i\Gamma} .
\end{align}
By combining these results, Eqs.~(\ref{eq:Lambdaproperty})-(\ref{eq:f3def}) can be derived.

\section{Derivation of Eq.~(\ref{eq:chiend})}
\label{app:analytic}

\begin{figure}[tb]
\centering
\includegraphics[width=80mm]{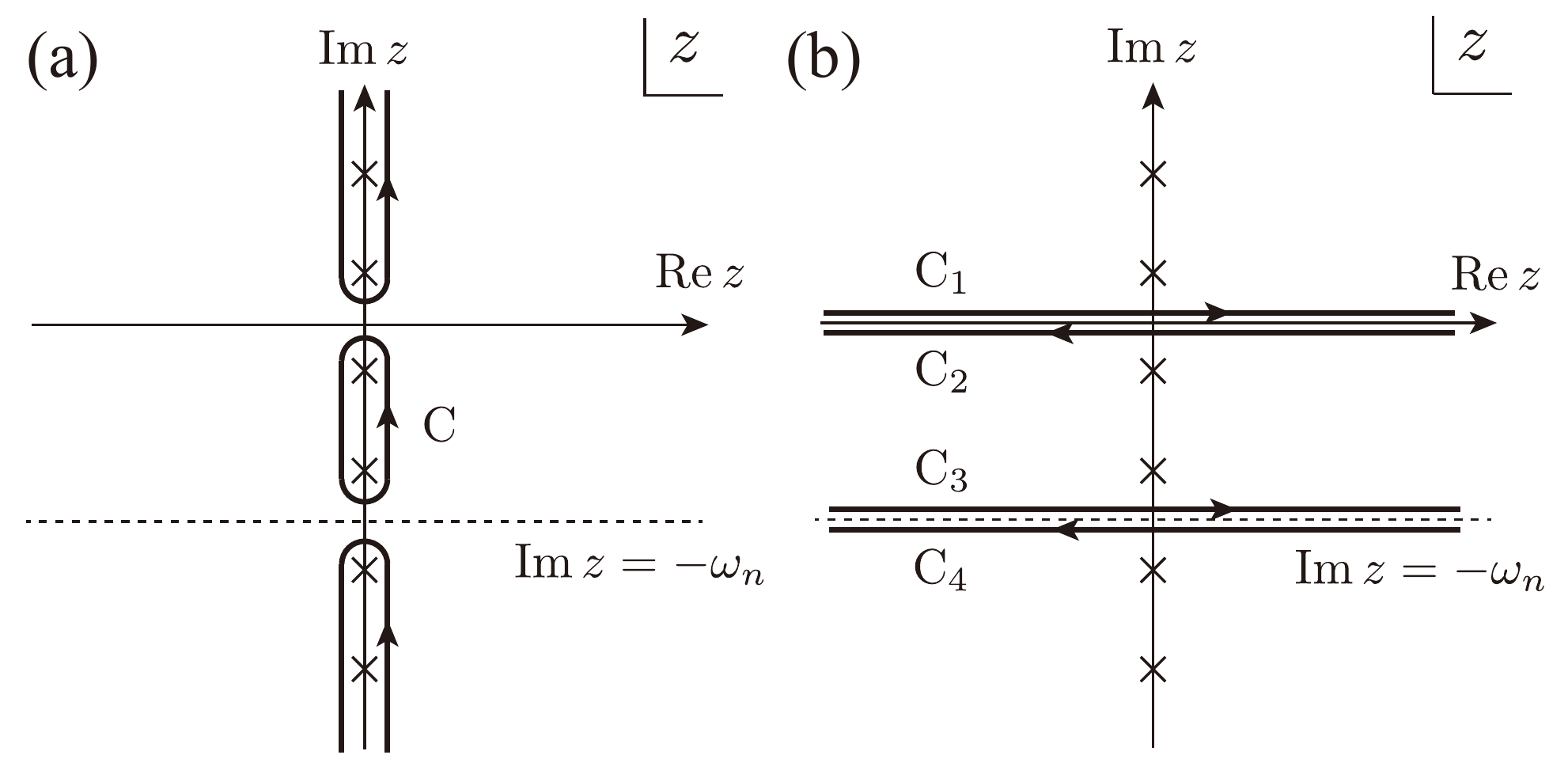}
\caption{Schematic picture of the change in the contour integral. (a) The original contour. (b) The modified contour.}
\label{fig:intcontour}
\end{figure}

In this Appendix, we give a detailed derivation of Eq.~(\ref{eq:chiend}) from Eq.~(\ref{eq:chibfranaly}).
First, we modify Eq.~(\ref{eq:chibfranaly}) as
\begin{align}
\chi({\bm 0},i\omega_n) 
&= \frac{1}{8\mathcal{A}} 
\sum_{{\bm k}}\sum_{\nu,\nu'}\Biggl{[}
\nu\nu'\sin 2(\phi-\theta) \, {\cal I}_{\nu\nu',1}\nonumber \\
&+\Bigl\{ 1-\nu\nu'\cos 2(\phi-\theta)\Bigl\} {\cal I}_{\nu\nu',2}\nonumber \\
&-i( 1-\nu\nu') {\cal I}_{\nu\nu',3}
\Biggl{]} , \label{eq:chibeforesum}
\end{align}
where
\begin{align}
{\cal I}_{\nu\nu',j}&\equiv \frac{1}{\tempinv}\sum_{i\omega_{m}}
\frac{X_{j}}{i\hbar \omega_m-E_{\bm{k}}^\nu+i\Gamma/2\,{\rm sgn}(\omega_m)} \nonumber \\
&\hspace{-5mm}  \times \frac{1}{i\hbar \omega_m+i\hbar \omega_n-E_{\bm{k}}^{\nu'}+i\Gamma/2\,{\rm sgn}(\omega_m+\omega_n)} ,
\end{align}
and $(X_1,X_2,X_3)=(X,Y,Z)$.
A standard procedure based on the residue integral enables us to express the sum 
${\cal I}_{\nu\nu',j}$ for $\omega_n>0$ as a complex integral on the contour ${\rm C}$ shown in Fig.~\ref{fig:intcontour}~(a).
This contour can be modified into a sum of the four contours, ${\rm C}_l$ ($l=1,2,3,4$), shown in Fig.~\ref{fig:intcontour}~(b).
Accordingly, 
${\cal I}_{\nu\nu',j}$ is written as
\begin{align}
{\cal I}_{\nu\nu',j} &=\sum_{l=1}^{4}
{\cal I}_{\nu\nu',j}^{{\rm C}_l}, \\
{\cal I}_{\nu\nu',j}^{{\rm C}_l}&=-\int_{{\rm C}_l}
\frac{dz}{2\pi i} 
\frac{f(z) X_{j}(z,i\omega_{n})}{z-E_{\bm k}^\nu + i\Gamma/2 \, {\rm sgn}({\rm Im}\, z)} \nonumber \\
&\times \frac{1}{z+i\hbar \omega_n-E_{\bm{k}}^{\nu'} +i\Gamma/2 \, {\rm sgn}({\rm Im}\,z+\omega_n)},
\end{align}
where $f(z)=1/(e^{\tempinv} z+1)$ is the Fermi distribution function.
The sum of the contributions from the two contours, ${\rm C}_2$ and ${\rm C}_3$, is calculated as
\begin{align}
& {\cal I}^{{\rm C}_{2}}_{\nu\nu',j}+{\cal I}^{C_{3}}_{\nu\nu',j} \nonumber \\
&= -\tilde{X}_{j}(i\omega_{n})\int \frac{dE}{2\pi i} f(E)\nonumber \\
&\hspace{4mm} \times \Biggl[- \frac{1}{E -E_{\bm{k}}^{\nu}-i\Gamma/2} 
\frac{1}{E+i\hbar \omega_n-E_{\bm{k}}^{\nu'}+i\Gamma/2} 
\nonumber \\
&\hspace{6mm} + \frac{1}{E -i\hbar \omega_n -E_{\bm{k}}^{\nu}-i\Gamma/2} \frac{1}{E-E_{\bm{k}}^{\nu'}+i\Gamma/2} \Biggr].
\label{eq:integ23}
\end{align}
Here, we have used the fact that $X_j(z,i\omega_n)$ is independent of $z$ for $0<{\rm Im}\, z<\omega_n$ from Eq.~(\ref{eq:Lambdaproperty}) and have defined its value as $\tilde{X}_j(i\omega_n)$ ($j=1,2,3$).
From Eqs.~(\ref{eq:X})-(\ref{eq:Z}), $\tilde{X}_j(i\omega_n)$ are calculated as
\begin{align}
& \tilde{X}_1(i\omega_n) = \frac{\tilde{\Lambda}_3(i\omega_n)}{(1-\tilde{\Lambda}_0(i\omega_n))^2-\tilde{\Lambda}_2(i\omega_n)^2{-\tilde{\Lambda}_3(i\omega_n)^2}}, \\
& \tilde{X}_2(i\omega_n) = {\frac{1-\tilde{\Lambda}_0(i\omega_n)-\tilde{\Lambda}_2(i\omega_n)}{(1-\tilde{\Lambda}_0(i\omega_n))^2-\tilde{\Lambda}_2(i\omega_n)^2{-\tilde{\Lambda}_3(i\omega_n)^2}}}, \\
& \tilde{X}_3(i\omega_n) = \frac{i}{1-\tilde{\Lambda}_0(i\omega_n)+\tilde{\Lambda}_1(i\omega_n)}.
\end{align}
By changing the integral variable to $E'=E-E_{\bm k}^{{\nu}}$ in the first term and to $E'=-(E-E_{\bm k}^{{\nu'}})$ in the second term in Eq.~(\ref{eq:integ23}), we obtain
\begin{align}
{\cal I}^{{\rm C}_{2}}_{\nu\nu',j}+{\cal I}^{{\rm C}_{3}}_{\nu\nu',j} &= -\tilde{X}_{j}(i\omega_{n})\int \frac{dE'}{2\pi i} 
\frac{1}{E'-i\Gamma/2} \nonumber \\
&\hspace{-7mm} \times \Biggl[ \frac{f(-E'+E_{\bm k}^{{\nu'}})-
f(E'+E_{\bm k}^{{\nu}})}
{E'+i\hbar \omega_n+E_{\bm k}^{{\nu}}-E_{\bm k}^{{\nu'}}+i\Gamma/2} \Biggr] .
\end{align}
Using formula (\ref{eq:ksumformula}), we replace the sum over ${\bm k}$ in Eq.~(\ref{eq:chibeforesum}) by the integral with respect to $\xi$ and $\varphi$.
We can perform the $\xi$-integral by using
\begin{align}
&-\int_{-\infty}^{\infty} d\xi \, [f(E'+E^{{\nu}}_{\bm k})-f(-E'+E^{{\nu'}}_{\bm k})] \nonumber \\
&\hspace{5mm} = 2E'+E^{{\nu}}_{\bm k}-E^{{\nu'}}_{\bm k}.
\end{align}
Then, by performing the $E'$-integral, we obtain
\begin{align}
&\int_{-\infty}^\infty d\xi\, ({\cal I}^{{\rm C}_{2}}_{\nu\nu',j}+{\cal I}^{{\rm C}_{3}}_{\nu\nu',j}) 
= \frac{i\hbar\omega_n \tilde{X}_{j}(i\omega_{n})}{E_{\bm k}^{\nu}-E_{\bm k}^{\nu'}+i\hbar\omega_n+i\Gamma}.
\end{align}
Next, let us consider the contribution from ${\rm C}_1$ and ${\rm C}_4$.
On these two contours, $X_j(z,i\omega_n)$ is independent of $z$ and its value is defined by  $\tilde{X}_j'(i\omega_n)$ ($j=1,2,3$).
Because $\Lambda_j(z,i\omega_n)$ ($j=0,1,2,3$) becomes zero for ${\rm Im}\, z<0$ or $\omega_n < {\rm Im}\, z$ from Eq.~(\ref{eq:Lambdaproperty}), $\tilde{X}_j'(i\omega_n)$ are given as
\begin{align}
\tilde{X}_1'(i\omega_n) = 0, \quad
\tilde{X}_2'(i\omega_n) = 1, \quad
\tilde{X}_3'(i\omega_n) = i. 
\end{align}
A similar calculation to that of ${\rm C}_2$ and ${\rm C}_3$ yields
\begin{align}
&\int_{-\infty}^\infty d\xi \, ({\cal I}^{{\rm C}_{1}}_{\nu\nu',j}+{\cal I}^{{\rm C}_{4}}_{\nu\nu',j}) 
=-\tilde{X}_{j}'(i\omega_n).
\end{align}
By substituting these results into Eq.~(\ref{eq:chibeforesum}), we obtain 
\begin{align}
\chi({\bm 0},i\omega_n) 
&=\frac{D(\epsilon_{\rm F})}{8} 
\sum_{\nu,\nu'}\int_0^{2\pi} \frac{d\varphi}{2\pi} \, \Biggl{[}
\nu\nu'\sin 2(\phi-\theta) \, \tilde{x}_1(i\omega_n) \nonumber  \\
&\hspace{5mm} +[1-\nu\nu'\cos 2(\phi-\theta)] [-1 + \tilde{x}_2(i\omega_n) ] \nonumber \\
&\hspace{5mm} -i(1-\nu\nu') [ -i + \tilde{x}_3(i\omega_n)] \Biggr].
\end{align}
where
\begin{align}
\tilde{x}_j(i\omega_n) = \frac{i\hbar\omega_n \tilde{X}_j(i\omega_n)}{E_{\bm k}^{\nu}-E_{\bm k}^{\nu'}+i\hbar\omega_n+i\Gamma}.
\end{align}
Finally, Eq.~(\ref{eq:chiend}) is derived by substituting the expressions for $\tilde{X}_j(i\omega_n)$ and by analytic continuation $i\omega_n \rightarrow \omega+i\delta$.

\section{Analytic Expression for $\beta/\alpha = 0$}
\label{app:alphazero}

In this appendix, we derive analytic expressions of the modulation of the Gilbert damping when $\beta/\alpha=0$, only the Rashba spin-orbit interaction exists, to see the quantitative effect of taking the vertex correction into account.
For $\beta/\alpha = 0$, the spin-splitting energy $2h_{\rm eff}=2\Delta_0$ is constant along the Fermi surface, and $\tilde{\Lambda}_j^R(\omega)$ ($j=0,1,2,3$) is simplified as
\begin{align}
\tilde{\Lambda}_{0}^R(\omega) &= \frac{i\Gamma}{4\Delta_0}\sum_{\nu\nu'}\frac{1}{\hbar\omega/\Delta_0+(\nu-\nu')+i\Gamma/\Delta_0},\label{eq:aob0L0}\\
\tilde{\Lambda}_{1}^R(\omega) &= \frac{i\Gamma}{4\Delta_0}\sum_{\nu\nu'}\frac{\nu\nu'}{\hbar\omega/\Delta_0+(\nu-\nu')+i\Gamma/\Delta_0},\label{eq:aob0L1}\\
\tilde{\Lambda}_{2}^R(\omega) &=
{\tilde{\Lambda}_{3}^R(\omega) }=0.
\end{align}
Then, we obtain the modulation of the Gilbert damping with the vertex corrections,
\begin{align}
\frac{\delta \alpha_{\rm G}}{\alpha_{{\rm G},0}} &\simeq 
\frac{\Delta_0}{2\pi\Gamma}
{\rm Re} \,
\Biggl{[}
\frac{\tilde{\Lambda}_0^R(\omega_0)}{1-\tilde{\Lambda}_0^R(\omega_0)}+ \frac{\tilde{\Lambda}_0^R(\omega_0)-\tilde{\Lambda}_1^R(\omega_0)}{1-\tilde{\Lambda}_0^R(\omega_0)+\tilde{\Lambda}_1^R(\omega_0)}
\Biggl{]}.\label{eq:vergil1}
\end{align}
The modulation of the Gilbert damping without the vertex correction is obtained by considering only the first-order term with respect to $\tilde{\Lambda}_j^R(\omega_0)$,
\begin{align}
\frac{\delta \alpha_{\rm G}^{\rm nv}}{\alpha_{{\rm G},0}} &\simeq 
\frac{\Delta_0}{2\pi\Gamma} {\rm Re} \,
\Biggl{[}
2\tilde{\Lambda}_0^R(\omega_0)-\tilde{\Lambda}_1^R(\omega_0)
\Biggl{]}.\label{eq:novergil1}
\end{align}
When $\Gamma \ll \Delta_0$, the contribution of $\nu=\nu'$ is dominant for the peak at $\omega_0=0$ and the modulation of the Gilbert damping can be analytically calculated as
\begin{align}
\frac{\delta \alpha_{\rm G}}{\alpha_{{\rm G},0} } &\simeq 
\frac{\Delta_0}{4\pi}\cdot\frac{\Gamma/2}{(\hbar\omega_0)^{2}+(\Gamma/2)^2},
\label{eq:ver-gil-omega0}\\
\frac{\delta \alpha_{\rm G}^{\rm nv}}{\alpha_{{\rm G},0}}
&\simeq  
\frac{\Delta_0}{4\pi} \cdot
\frac{\Gamma}{(\hbar\omega_0)^{2}+\Gamma^{2}}.
\label{eq:nover-gil-omega0}
\end{align}
This indicates that the peak width is halved by taking the vertex correction into account, 
which is consistent with the results shown in Figs.~\ref{fig:2Dplot}~(b) and \ref{fig:2Dplot}~(c).

\begin{figure*}[tb]
\centering
\includegraphics[width=170mm]{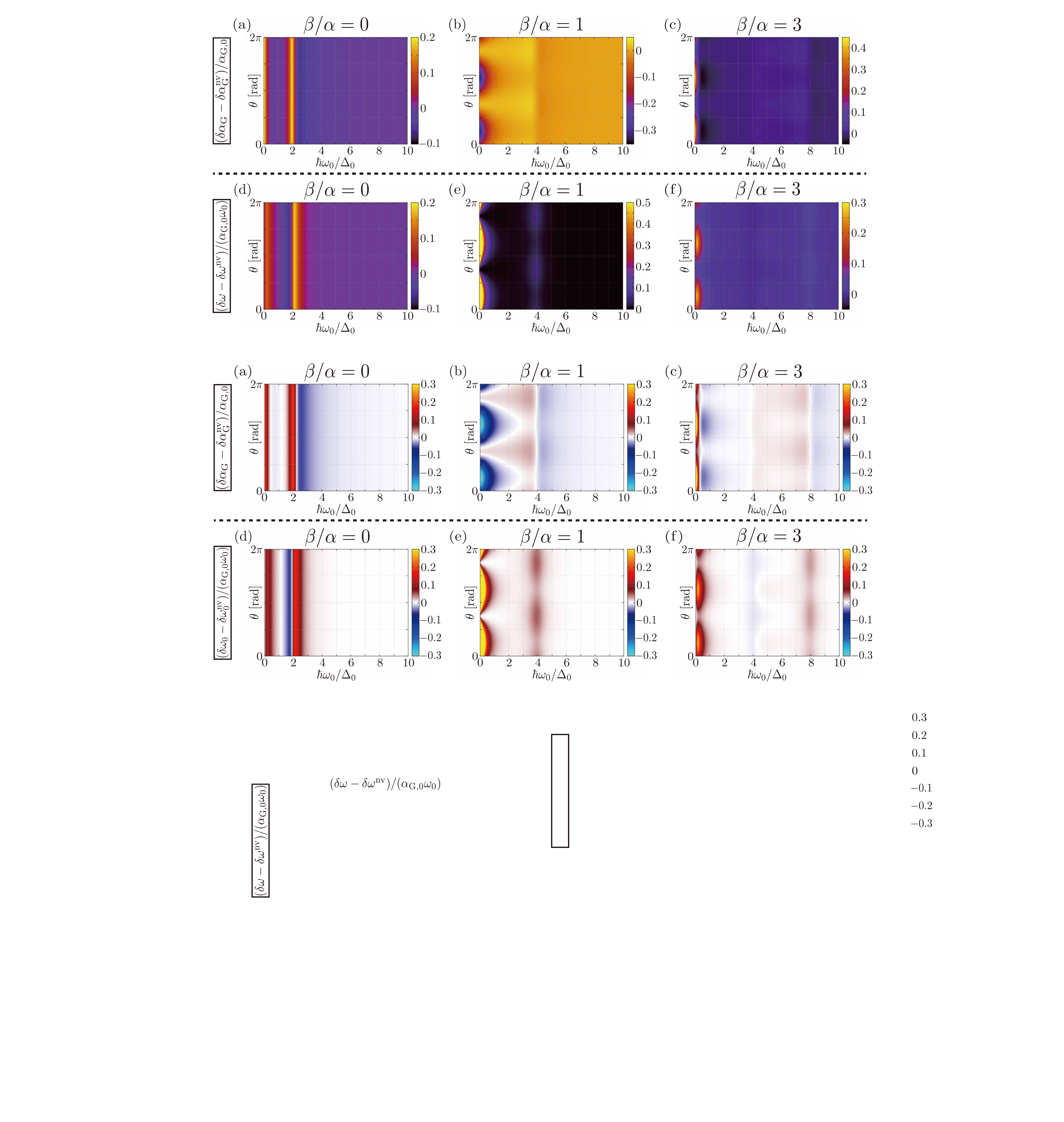}
\caption{(Upper panels) 
Change of the Gilbert damping due to the vertex correction, $\delta\alpha_{\rm G}-\delta\alpha^{\rm nv}_{\rm G}$, for (a) $\beta/\alpha=0$, (b) $\beta/\alpha=1$, and (c) $\beta/\alpha=3$.
(Lower panels) 
Change of the FMR frequency due to the vertex correction, 
$\delta\omega_0-\delta\omega^{\rm nv}_0$,
for (d) $\beta/\alpha=0$, (e) $\beta/\alpha=1$, and (f) $\beta/\alpha=3$.
The horizontal axes are the FMR frequency, $\omega_0=\gamma h_{\rm dc}$, whereas the vertical axes are the azimuth angle of the spontaneous spin polarization, $\theta$, in the FI. 
In all the plots, we have set $\Gamma/\Delta_{0}=0.5$.
}
\label{fig:densitySIGN}
\end{figure*}

In a similar way, we can evaluate the modulation of the Gilbert damping near the peak at $\omega_0 = 2\Delta_{0}/\hbar$ as
\begin{align}
\frac{\delta \alpha_{\rm G}}{\alpha_{{\rm G},0}} &\simeq 
 \frac{\Delta_0}{4\pi}\cdot\biggl{[}
\frac{1}{2}\frac{3\Gamma/4}{(\hbar\omega_0-2\Delta_0)^{2}+(3\Gamma/4)^{2}}
\nonumber \\
&\hspace{8mm}  + \frac{\Gamma/2}{(\hbar\omega_0-2\Delta_0)^{2}+(\Gamma/2)^{2}}
\biggl{]},\label{eq:ver-gil-omega2h}\\
\frac{\delta \alpha_{\rm G}^{\rm nv}}{\alpha_{{\rm G},0}} &\simeq
\frac{\Delta_0}{4\pi} \cdot
\frac{3\Gamma/2}{(\hbar\omega_0-2\Delta_0)^{2}+\Gamma^{2}} . \label{eq:nover-gil-omega2h}
\end{align}
As well, for the peak at $\omega_0=2\Delta_0/\hbar$, the peak width becomes smaller when the vertex correction is taken into account.
This observation is consistent with the results shown in Figs.~\ref{fig:2Dplot}~(b) and \ref{fig:2Dplot}~(c).
For a finite value of $\Gamma$, a sum of Eqs.~(\ref{eq:ver-gil-omega0}) and (\ref{eq:ver-gil-omega2h}) [Eqs.~(\ref{eq:nover-gil-omega0}) and (\ref{eq:nover-gil-omega2h})] gives a better analytic form which fits the numerical result with (without) the vertex correction. 
Note that $\delta\alpha_{\rm G}$ and $\delta\alpha^{\rm nv}_{\rm G}$ depend on the impurity potential strength, $u$, and impurity concentration,
$n_i$, through $\Gamma=2\pi n_i u^2 D(\epsilon_{\rm F})$ [see Eq.~(\ref{def:Gamma})]. 
As shown in Eqs.~(\ref{eq:ver-gil-omega0})-(\ref{eq:nover-gil-omega2h}),
the peak widths of the Lorentzian functions in $\delta\alpha_{\rm G}$ and $\delta\alpha^{\rm nv}_{\rm G}$ are determined by $\Gamma$ [see Figs.~\ref{fig:2Dplot}~(b) and \ref{fig:2Dplot}~(c)].
It is remarkable that the peak width in $\delta \alpha_{\rm G}$ is reduced from $\Gamma$ to $\Gamma/2$ by taking the vertex correction into account.
To summarize the effect of the vertex correction, we show $\delta\alpha_{\rm G}-\delta\alpha^{\rm nv}_{\rm G}$ and $\delta\omega_0-\delta\omega^{\rm nv}_0$
in Figs.~\ref{fig:densitySIGN}~(a) and \ref{fig:densitySIGN}~(d), respectively.
We find that the vertex correction modifies mainly the peak width around $\hbar \omega_0/\Delta_0 = 0$ and 2, in consistent with the above analytic expressions.

Finally, we note that the same analytical expressions for $\delta \alpha_{\rm G}$ and $\delta \alpha_{\rm G}^{\rm nv}$ can be obtained for the case of $\alpha/\beta=0$, i.e., when only the Dresselhaus spin-orbit interaction exists.
We also note that for general values of $\beta/\alpha$, $\delta \alpha_{\rm G}$ and $\delta \alpha_{\rm G}^{\rm nv}$ depend on $\Gamma$ in a more complicated way.

\section{Analytic Expression for $\beta/\alpha = 1$}
\label{app:alphaone}

In this Appendix, we derive analytic expressions of the modulation of the Gilbert damping when $\beta/\alpha=1$.
In this case, the effective Zeeman field is parallel to the $(-1,1,0)$ direction and its amplitude is given as
\begin{align}
h_{\rm eff}(\varphi) &= 2\Delta_0 |\sin(\varphi + \pi/4)|.
\end{align}
Then, $\tilde{\Lambda}_j^R(\omega)$ ($j=0,1,2,3$) becomes
\begin{align}
\tilde{\Lambda}_{0}^R(\omega) &= \frac{i\Gamma}{4\Delta_0}\sum_{\nu\nu'}{\cal J}_{\nu\nu'} \\
\tilde{\Lambda}_{1}^R(\omega) &= \frac{i\Gamma}{4\Delta_0}\sum_{\nu\nu'}\nu\nu' {\cal J}_{\nu\nu'}, \\
\tilde{\Lambda}_{2}^R(\omega) &= -\sin2\theta\, \tilde{\Lambda}_{1}^R(\omega) ,\\
{\tilde{\Lambda}_{3}^R(\omega) }&= -\cos2\theta\, \tilde{\Lambda}_{1}^R(\omega) 
\end{align}
where
\begin{align}
{\cal J}_{\nu\nu'}(\omega) &\equiv\int_{0}^{2\pi}\frac{d\varphi}{2\pi}\frac{\Delta_0}{\hbar\omega+(\nu-\nu')h_{\rm eff}(\varphi)+i\Gamma}.
\end{align}

In the case of $\theta=\pi/4$, the modulation of the Gilbert damping with the vertex correction is expressed as
\begin{align}
\frac{\delta\alpha_{\rm G}}{\alpha_{{\rm G},0}} =& \frac{\Delta_0}{2\pi\Gamma}\, {\rm Re} \,\Biggl{[} -2
+
\frac{1}{1-\tilde{\Lambda}_0^R(\omega_0)+\tilde{\Lambda}_{1}^R(\omega_0)}\nonumber \\
&\hspace{10mm} +
\frac{1}{1-\tilde{\Lambda}_0^R(\omega_0)-\tilde{\Lambda}_{1}^R(\omega_0)}
\Biggl{]}.
\label{eq:aob1vergilapp}
\end{align}
The third term of the above equation is calculated as
\begin{align}
\frac{1}{1-\tilde{\Lambda}_0^R(\omega_0)-\tilde{\Lambda}_{1}^R(\omega_0)}
&=\frac{1}{1-\frac{i\Gamma}{\hbar \omega_0+i\Gamma}}
=\frac{\hbar\omega_0+i\Gamma}{\hbar\omega_0}.
\end{align}
This indicates that the expansion with respect to $\tilde{\Lambda}_j^R$ can not be allowed for $\omega_0 \ll \Gamma$.
This is why the modulation without the vertex correction, which is obtained by taking from the first-order term of $\tilde{\Lambda}_{j}^R$ in Eq.~(\ref{eq:aob1vergilapp}) as
\begin{align}
\frac{\delta \alpha^{\rm nv}_{\rm G}}{\alpha_{{\rm G},0}}
&= \frac{\Delta_0}{2\pi\Gamma}
{\rm Re} \, 
\biggl{[} 2\tilde{\Lambda}_{0}^R(\omega_0) \biggl{]},
\end{align}
gives a different result near $\omega_0\simeq0$.
Actually, for $\theta=\pi/4$, $\delta \alpha_{\rm G}$ and $\delta \alpha^{\rm nv}_{\rm G}$ are calculated as
\begin{align}
\frac{\delta\alpha_{\rm G}}{\alpha_{{\rm G},0}}
&= \frac{\Delta_0}{2\pi\Gamma} {\rm Re} \,
\Biggl{[} \frac{i\frac{\Gamma}{2\Delta_0}({\cal J}_{+-}+{\cal J}_{-+})}{1-i\frac{\Gamma}{2\Delta_0}({\cal J}_{+-}+{\cal J}_{-+})} \Biggl{]}, 
\label{app:double}
\\
\frac{\delta\alpha^{\rm nv}_{\rm G}}{\alpha_{{\rm G},0}}
&= \frac{1}{4\pi} {\rm Re} \,
\Biggl{[}i\biggl({\cal J}_{+-} + {\cal J}_{-+} + {\cal J}_{++} + {\cal J}_{--}\biggr)
\Biggr{]}.
\end{align}
Note that Eq.(\ref{app:double}) is not valid for $\omega_0=0$.
As indicated from the absence of ${\cal J}_{++}$ and ${\cal J}_{--}$, the graph of $\delta\alpha_{\rm G}(\omega_0)$ has no peak at zero frequency even though $\delta\alpha_{\rm G}^{\rm nv}(\omega_0)$ has a peak there.
This observation is consistent with  Figs.~\ref{fig:2Dplot}~(e) and \ref{fig:2Dplot}~(f).

In the case of $\theta=-\pi/4$, the modulations of the Gilbert damping with and without the vertex correction are 
\begin{align}
\frac{\delta\alpha_{\rm G}}{\alpha_{{\rm G},0}}
&= \frac{\Delta_0}{\pi\Gamma} {\rm Re} \,
\Biggl{[} \frac{i\frac{\Gamma}{2\Delta_0}({\cal J}_{+-}+{\cal J}_{-+})}{1-i\frac{\Gamma}{2\Delta_0}({\cal J}_{+-}+{\cal J}_{-+})}\Biggl{]}, \label{piover4tmp}
\\
\frac{\delta\alpha^{\rm nv}_{\rm G}}{\alpha_{{\rm G},0}}
&= \frac{1}{2\pi} {\rm Re} \,
\Biggl[i({\cal J}_{+-}+{\cal J}_{-+})\Biggr] .
\end{align}
Note that $\delta\alpha^{\rm nv}_{\rm G}$ is obtained by taking the first-order term in Eq.~(\ref{piover4tmp}).
As indicated by the absence of the terms, ${\cal J}_{++}$ and ${\cal J}_{--}$, neither $\delta\alpha_{\rm G}$ nor $\delta\alpha_{\rm G}^{\rm nv}$ has any structure around $\omega_0=0$.
It can be checked that these two expressions give almost the same result when $\Gamma \lesssim \Delta_0$, which is consistent with Figs.~\ref{fig:2Dplot}~(e) and \ref{fig:2Dplot}~(f).
Note as well that $\delta \alpha_{\rm G}$ is just doubled compared with the result for $\theta=\pi/4$ in Eq.~(\ref{app:double}).

To summarize the effect of the vertex correction, we show $\delta\alpha_{\rm G}-\delta\alpha^{\rm nv}_{\rm G}$ and $\delta\omega_0-\delta\omega^{\rm nv}_0$ in Figs.~\ref{fig:densitySIGN}~(b) and \ref{fig:densitySIGN}~(e), respectively.
We find that the vertex correction modifies mainly the peak width around $\hbar \omega_0/\Delta_0 = 0$.
In addition, the broad peak in the range of $0 < \hbar\omega_0 < 2\Delta_0$ is enhanced or suppressed depending on the azimuth angle of the ordered spin. 
These features are consistent with the above analytic expressions.
We note that similar features are observed for $\beta/\alpha = 3$ as seen in Figs.~\ref{fig:densitySIGN}~(c) and \ref{fig:densitySIGN}~(f).

\section{Approximate Expressions near $\beta/\alpha = 1$}
\label{app:nearone}

In this Appendix, we derive the approximate expressions Eqs.~(\ref{eq:Approx1}) and (\ref{eq:Approx2}) for $\beta/\alpha = 1+\delta$ ($\delta \ll 1$) and $\omega \simeq 0$.
For $\beta/\alpha = 1+\delta$ ($\delta \ll 1$), we can use the approximation,
\begin{align}
\cos2(\phi-\theta)
&\simeq \sin2\theta\Bigl{(}-1
+\frac{(h_{x}+h_{y})^{2}}{h_{\rm eff}^{2}}\Bigl{)},\\
\sin2(\phi-\theta)&\simeq \cos2\theta\Bigl{(}-1
+\frac{(h_{x}+h_{y})^{2}}{h_{\rm eff}^{2}}\Bigl{)}.
\end{align}
Then, we obtain
\begin{align}
\tilde{\Lambda}^{R}_2 &\simeq X \sin2\theta, \\
\tilde{\Lambda}^{R}_3 &\simeq X \cos2\theta,\\
X &\equiv \frac{i\Gamma}{4}\int_{0}^{2\pi}\frac{d\varphi}{2\pi} \sum_{\nu\nu'}\frac{\nu \nu'\Bigl{(}-1+\frac{(h_{x}+h_{y})^{2}}{h_{\rm eff}^{2}}\Bigl{)}}{\hbar\omega+(\nu-\nu')h_{\rm eff}+i\Gamma} 
\label{Xdef}
\end{align}
in the low-frequency region.
Here, the contribution of the second term of the bracket in Eq.~(\ref{Fomegadef}) does not have a singularity at $\omega_0 = 0$ because $\tilde{\Lambda}_0^R$ and $\tilde{\Lambda}_1^R$ do not depend on the effective Zeeman field ${\bm h}_{\rm eff}$.
Therefore, the singularity comes from the first term of the bracket in Eq.~(\ref{Fomegadef}) and we can approximate $F(\omega)$ as
\begin{align}
F(\omega) & \simeq \frac{\Delta_0}{2\pi i\Gamma}
\frac{\tilde{\Lambda}^{R}_0(1-\tilde{\Lambda}^{R}_0)-\tilde{\Lambda}^{R}_2(1-\tilde{\Lambda}^{R}_2)+(\tilde{\Lambda}^{R}_3)^{2}}{(1-\tilde{\Lambda}^{R}_0)^2-(\tilde{\Lambda}^{R}_2)^2{-(\tilde{\Lambda}^{R}_3)^2}} \nonumber \\
&= \frac{\Delta_0}{2\pi i\Gamma} \left[ -1 + \frac{(1-\sin 2\theta)/2}{1-\tilde{\Lambda}_0^R-X}
+ \frac{(1+\sin 2\theta)/2}{1-\tilde{\Lambda}_0^R+X} \right].
\label{appDeq}
\end{align}
Finally, using the equation,
\begin{align}
1-\tilde{\Lambda}_0^R+X 
= \frac{\Gamma_s}{\Gamma} - i \frac{\hbar \omega}{\Gamma}
+ {\cal O}(\omega^2),
\label{appDeq2}
\end{align}
we find that the third term in the bracket in Eq.~(\ref{appDeq}) is divergent at $\omega=0$ in the limit of $\delta \rightarrow 0$ since the denominator vanishes.
By substituting Eq.~(\ref{appDeq2}) into Eq.~(\ref{appDeq}), the most singular part is calculated as
\begin{align}
F(\omega) \simeq \frac{\Delta_0}{2\pi i} \frac{\sin^2(\theta+\pi/4)}{\Gamma_s + i \hbar \omega}.
\end{align}
Using Eqs.~(\ref{eq:resltome}) and (\ref{eq:resltgil}), it is straightforward to obtain Eqs.(\ref{eq:Approx1}) and (\ref{eq:Approx2}).

\bibliography{ref}

\begin{thebibliography}{74}
\bibitem [{foo()}]{footnoteapp}%
  \BibitemOpen
  \href@noop {} {}\bibinfo {howpublished} {Although the interfacial exchange
  coupling in the present system has not been examined experimentally, its
  order can be estimated from the exchange bias between a ferromagnetic
  insulator and a metal~\cite{NOGUES1999203}. This indicates that the
  interfacial exchange coupling for a YIG/semiconductor magnetic junction is
  much less than $0.1 \, {\rm meV}$, justifying our approximation.}\BibitemShut
  {Stop}%
\end{thebibliography}

\end{document}